\documentclass[aps,prc,twocolumn,superscriptaddress,showkeys,nofootinbib]{revtex4-1}

\usepackage[utf8]{inputenc}

\usepackage{graphicx,color}
\usepackage{amsmath,amssymb,amsfonts}
\usepackage{xspace}	% Include xspace
\usepackage{ulem}

\usepackage{hyperref}

\newcommand{\softsh}[1]{\texttt{#1}}
\newcommand{\soft}[1]{\texttt{#1}\xspace}
\newcommand{\beagle}{\soft{BeAGLE}}

\newcommand{\liseppcutesh}{\softsh{LISE$^{++}_{cute}$}}
\newcommand{\inclliege}{\soft{INCL-Li\`ege}}

\usepackage{lineno}
\begin{document}

\title{
Producing and Studying Rare Isotopes in $e{+}A$ Collisions at the Electron–Ion Collider
}
%------------------------------------------------------------------
\author{Mark Ddamulira}
\affiliation{Department of Physics, Texas Southern University, Houston, TX 77004} 
\affiliation{Department of Physics and Astronomy, Michigan State University, East Lansing, Michigan 48824}
%------------------------------------------------------------------
\author{Abhay Deshpande}
\affiliation{Center for Frontiers in Nuclear Science (CFNS), Dept. of Physics and Astronomy, Stony Brook University, Stony Brook, NY 11794-3800} 
\affiliation{Department of Physics, Brookhaven National Laboratory, Upton, NY 11973-5000}
%------------------------------------------------------------------
\author{Mark C. Harvey}
\email{mark.harvey@tsu.edu}
\affiliation{Department of Physics, Texas Southern University, Houston, TX 77004}
%------------------------------------------------------------------
\author{Wenliang Li} 
\affiliation{Department of Physics and Astronomy, Mississippi State University, Starkville, MS 39762}
%------------------------------------------------------------------
\author{Niseem Magdy} 
%\email{niseem.abdelrahman@tsu.edu}
\thanks{Corresponding author: niseem.abdelrahman@tsu.edu}
\affiliation{Department of Physics, Texas Southern University, Houston, TX 77004} 
\affiliation{Department of Physics, Brookhaven National Laboratory, Upton, NY 11973-5000}

%------------------------------------------------------------------
\author{Brynna Moran} 
\affiliation{Center for Frontiers in Nuclear Science (CFNS), Dept. of Physics and Astronomy, Stony Brook University, Stony Brook, NY 11794-3800} 
%------------------------------------------------------------------
\author{Pawel Nadel-Turonski} 
\affiliation{Department of Physics and Astronomy, University of South Carolina, Columbia, SC 29208}
%------------------------------------------------------------------
\author{Charles Joseph Naim} 
\affiliation{Center for Frontiers in Nuclear Science (CFNS), Dept. of Physics and Astronomy, Stony Brook University, Stony Brook, NY 11794-3800} 
%------------------------------------------------------------------
\author{Stacyann Nelson} 
\affiliation{Department of Physics, Morgan State University, Baltimore, MD 21251}
%------------------------------------------------------------------
\author{Isaiah Richardson} 
\affiliation{Department of Physics and Astronomy, Michigan State University, East Lansing, Michigan 48824}
\affiliation{Facility for Rare Isotope Beams, Michigan State University, East Lansing, MI 48824}
%------------------------------------------------------------------
\author{Barak A. Schmookler}
\email{baschmoo@central.uh.edu}
\affiliation{Department of Physics, University of Houston, Houston, TX 77204-5005}
%------------------------------------------------------------------
\author{Oleg B. Tarasov} 
\affiliation{Facility for Rare Isotope Beams, Michigan State University, East Lansing, MI 48824}
%------------------------------------------------------------------

%------------------------------------------------------------------

%------------------------------------------------------------------
\medskip
%------------------------------------------------------------------
%------------------------------------------------------------------

%------------------------------------------------------------------
%------------------------------------------------------------------
%\date{\today}

%------------------------------------------------------------------
\begin{abstract}
The Electron--Ion Collider (EIC) offers a unique environment to study kinematically controlled lepton--nucleus ($e{+}A$) reactions, where a primary hard scattering is followed by an intranuclear cascade and the subsequent statistical de-excitation of the nuclear remnant. Utilizing the \soft{BeAGLE} model, we demonstrate that event-by-event fluctuations in nucleon removal and energy deposition populate a diverse ensemble of excited remnants. Furthermore, we show that varying the target mass systematically shifts the distribution of these remnants across the $(N, Z)$ plane. Although this excited prefragment remnant is not directly observable, its properties are shown to be strongly correlated with final-state fragments; specifically, the largest nuclear residue and the intensity of evaporation yield serve as effective experimental proxies for event-level remnant characterization. We also evaluate photon observables essential for nuclear spectroscopy. While various photon sources overlap significantly in pseudorapidity, we find that in the nucleus-rest frame, the low-energy spectrum is dominated by de-excitation $\gamma$ rays and exhibits distinct discrete structures. These findings motivate an EIC research program that correlates rare-isotope production and de-excitation radiation with well-defined initial conditions, providing a collider-based approach to nuclear spectroscopy that is complementary to existing fixed-target facilities.
\end{abstract}
%----------------------------------------------------------------------
%----------------------------------------------------------------------
\maketitle
%----------------------------------------------------------------------
%\linenumbers
%----------------------------------------------------------------------
%----------------------------------------------------------------------
\section*{Introduction}
\begin{figure*}[htbp]
\centering{
\includegraphics[width=1.0\linewidth, height=8.5cm]{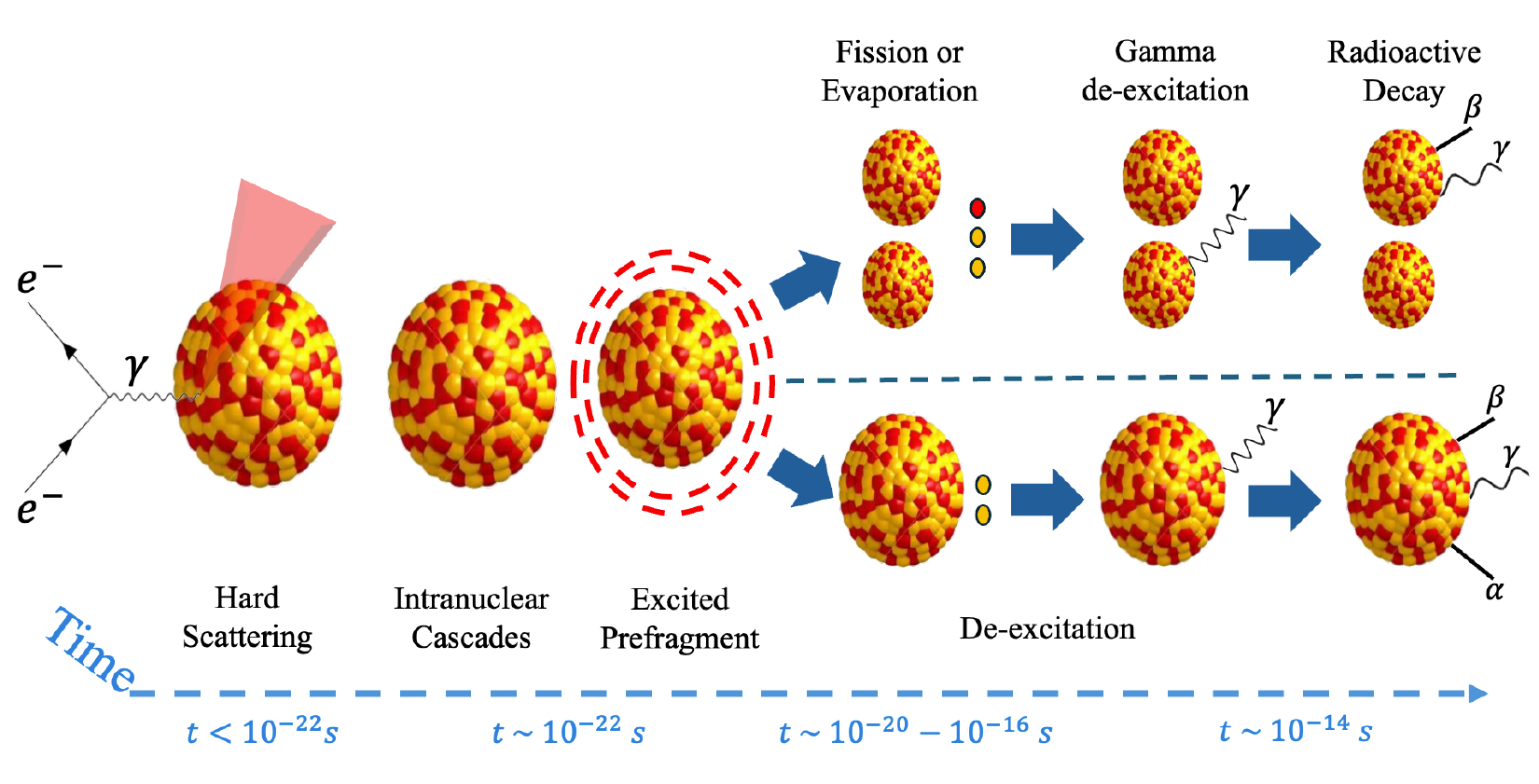}
\vskip -0.4cm
\caption{
Schematic illustration of the primary reaction stages in electron–nucleus ($e+A$) collisions. The high-energy virtual photon carries an energy $\nu$ ranging from $0.05$ to $17~\mathrm{GeV}$, corresponding to electron beam energies of $E_e = 5\text{--}18~\mathrm{GeV}$~\cite{Accardi:2012qut}. This virtual photon initiates a hard scattering on a quark, followed by parton showering and hadronization, intranuclear cascade and pre-equilibrium dynamics, and finally nuclear de-excitation via particle evaporation and $\gamma$ emission from the excited residual nucleus, followed by the characteristic decay of the resulting radioactive isotope.
\label{fig:fig1}
}
}
\vskip -0.1cm
\end{figure*}

%----------------------------------------------------------------------
The upcoming Electron--Ion Collider (EIC)~\cite{AbdulKhalek:2021gbh, Accardi:2012qut} will enable systematic studies of how nuclei are excited and subsequently de-excite in $e+A$ collisions under collider kinematics. In the nuclear rest frame, the reaction proceeds through three broad stages (Fig.~\ref{fig:fig1}): (i) an instantaneous hard scattering~\cite{Piller:1995kh, Arneodo:1994wf}, (ii) a rapid intranuclear cascade (INC) on a timescale of $\sim 10^{-22}$~s~\cite{Cugnon:1982qw}, and (iii) nuclear de-excitation over $\sim 10^{-20}$--$10^{-16}$~s~\cite{Weisskopf:1937zz, Charity:2010, Bondorf:1995ua}. Relative to fixed-target measurements~\cite{Adderley:2024czm}, collider boosts help separate these stages in the laboratory frame while the scattered lepton provides direct access to the initial-state kinematics driving the nuclear response~\cite{Magdy:2025cse}.
%-------------------------------------------------------------

%-------------------------------------------------------------
%-------------------------------------------------------------

%-------------------------------------------------------------
As illustrated in Fig.~\ref{fig:fig1}, the excited nuclear remnant emerging at the end of the cascade stage is a pre-equilibrium system shaped by (a) nucleon removal and the recoil imparted to the remnant, (b) partonic energy loss to the nuclear medium, and (c) secondary hadronic interactions within the nucleus. It is characterized by an excitation energy $E^{\ast}$ and by a mass number ($A^{\ast}$), proton number ($Z^{\ast}$), and neutron number ($N^{\ast}$) that generally differ from those of the target nucleus ($A$, $Z$, and $N$). The deviation $A-A^{\ast}$ depends on (i) the beam conditions (e.g., energy and target size) and (ii) the dynamics of the earlier reaction stages, leading to a broad distribution of accessible $A^{\ast}$ values. The subsequent de-excitation of this remnant can proceed through two limiting pathways: Case-1, cooling dominated by particle emission and fragmentation, producing a single heavy residue ($A^{\prime}$) accompanied by emitted hadrons (e.g., nucleons and light clusters); and Case-2, a binary breakup (fission) into two large fragments ($A^{\prime}$ and $A^{\prime\prime}$) plus hadrons. In the EIC environment, where the scattered lepton tags the hard interaction, far-forward detectors can measure the surviving nuclear fragments, these final de-excitation products (including $\gamma$ rays from the excited residue) can be studied in correlation with the underlying prefragment phase space $(A^{\ast},Z^{\ast},E^{\ast})$, enabling a differential spectroscopy program anchored to well-defined initial conditions. In this way, the EIC can extend and complement current rare-isotope programs~\cite{Crawford:2023txq}.
%-------------------------------------------------------------

%------------------------------------------------------------
Rare isotopes are atomic nuclei with typical neutron-to-proton ratios that lie far from the valley of $\beta$ stability. Characterized by short half-lives and low binding energies, these nuclides provide stringent tests for nuclear forces, the evolution of shell structure, and the emergence of collective phenomena away from stability. These nuclides also supply critical inputs to astrophysical models of element formation~\cite{Horowitz:2018ndv}. In this context, experimental measurements of masses, half-lives, and decay properties are necessary to constrain the pathways that set final abundances and freeze-out dynamics during the r-process.

\begin{figure}[htb]
\centering{
\includegraphics[width=1.0\linewidth]{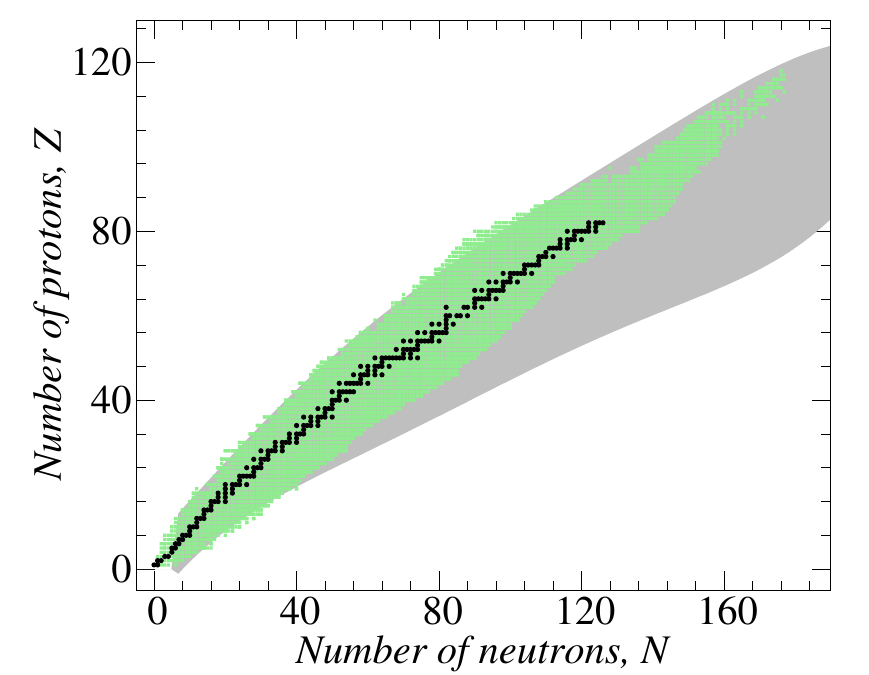}
\vskip -0.2cm
\caption{
Chart of nuclides in the $(N, Z)$ plane. Black squares indicate stable isotopes defining the valley of stability. The green region represents experimentally observed isotopes, while the gray area illustrates the predicted range of bound nuclei extending out to the theoretical drip lines. The presented data are from Refs.~\cite{NNDC_NuDat, Neufcourt:2020nme}.
\label{fig:fig0}
}
}
\vskip -0.1cm
\end{figure}

The current landscape of known nuclides is summarized in the $(N, Z)$ plane shown in Fig.~\ref{fig:fig0}~\cite{NNDC_NuDat, Neufcourt:2020nme}, which serves as the foundational map for rare-isotope science. Over the past several decades~\cite{Crawford:2023txq}, advancements in accelerators, targets, ion sources, separators, and detector systems have dramatically expanded experimental reach. As a result, the chart of nuclides has evolved from a sparse set of stable species into a dense network of measured isotopes. This progress is supported by curated nuclear-data services, such as NNDC/ENSDF and the IAEA LiveChart, which provide essential references for discovery and precision metrology~\cite{NNDC_NuDat, IAEA_LiveChart}. As illustrated in Fig.~\ref{fig:fig0}, the black squares represent stable isotopes defining the valley of stability, while the green region marks experimentally observed unstable nuclides. The surrounding gray band indicates the theoretically predicted domain of bound nuclei extending toward the drip lines; the continuing effort to populate this gray region with experimental measurements remains at the heart of modern rare-isotope research.
%------------------------------------------------------------

%------------------------------------------------------------
%------------------------------------------------------------

%------------------------------------------------------------
This experimental progress is underpinned by a global network of dedicated laboratories. The Facility for Rare Isotope Beams (FRIB) at Michigan State University serves as a next-generation heavy-ion accelerator and user facility, delivering fast, stopped, and reaccelerated beams across a vast expanse of the nuclide chart~\cite{FRIB, DOEFRIB, FRIB_OneYear}.
Complementing FRIB are several other flagship facilities: the RI Beam Factory (RIBF) at RIKEN~\cite{RIBF}, ISOLDE at CERN~\cite{ISOLDE, ISOLDEsetups}, GANIL/SPIRAL2~\cite{GANILSPIRAL2}, and the FAIR Super-FRS/NUSTAR program at GSI~\cite{FAIR, FAIR_NUSTAR}. Together, these institutions provide a powerful combination of high-intensity in-flight fragmentation, Isotope-Separation On-Line (ISOL) production, and high-resolution separation of reaction products.
Furthermore, capabilities at TRIUMF's ISAC/ARIEL complex~\cite{TRIUMF_ARIEL, TRIUMF_TIGRESS}, Argonne's ATLAS facility with ($\nu$)CARIBU~\cite{ATLAS_CARIBU, nuCARIBU}, and the IGISOL/JYFL facility~\cite{JYFL_IGISOL} broaden the global portfolio of available beam species, energy regimes, and experimental techniques. Collectively, these laboratories define the current frontier of nuclear structure and nuclear astrophysics, characterized by high intensity, expansive isotopic reach, and detection systems optimized for the most remote regions of the nuclear landscape.
%------------------------------------------------------------

%------------------------------------------------------------
At present, no operating collider is dedicated specifically to the production and investigation of rare isotopes. While heavy-ion colliders such as RHIC and the LHC have observed photon-induced processes in ultraperipheral collisions providing clean demonstrations of electromagnetic dissociation and photonuclear reactions, these measurements are not optimized for comprehensive isotope-production mapping or detailed spectroscopy across the nuclide chart~\cite{UPC_Pb208, BNL_RHICvsEIC}. This leaves a significant opportunity for a complementary collider-based approach, where complex nuclear remnants and their de-excitation signatures can be studied under precisely controlled initial conditions. Conceptually related ideas have recently emerged within the EIC community, suggesting that the EIC can serve not only as a probe of partonic structure but also as a laboratory for rare-isotope science through the controlled fragmentation of the nuclear beam~\cite{AbdulKhalek:2021gbh, CORE:2022rso, Bertulani:2024mqe, Kim:2026ytt}.
%------------------------------------------------------------

%------------------------------------------------------------
%------------------------------------------------------------
%------------------------------------------------------------
\section*{Motivation}
%------------------------------------------------------------
The central question, however, is how such a program would complement fixed-target rare-isotope measurements rather than duplicate them. In many fixed-target production experiments, the emphasis is necessarily on inclusive yields integrated over the (often imperfectly known) excitation and energy-loss history through material~\cite{Taieb:2003jj, Bernas:2003hd}, whereas the EIC can access differential, lepton-tagged correlations that connect fragment production and de-excitation observables directly to well-defined initial-state kinematics and to constrained regions of the prefragment phase space $(A^{\ast}, Z^{\ast}, E^{\ast})$. The EIC provides several qualitative advantages that directly map onto the observables pursued in this work:\\
(i) \emph{multi-species systematics at fixed beam energy and fixed forward acceptance:} the collider can operate with multiple ion species at the same beam energy, enabling measurements across a broad span of $(A, Z)$ without changing target thickness, magnetic rigidity settings, or detector geometry. This enables controlled systematics of prefragments and final fragments within a single experimental configuration, providing stringent tests of cascade and de-excitation modeling over a wide range of nuclei.\\
(ii) \emph{model disentanglement via lepton-tagging and prefragment constraints:} by correlating the scattered lepton kinematics with the observed fragment(s), the EIC provides differential access to the underlying excitation process. Tagging regions of the prefragment phase space $(A^{\ast}, Z^{\ast}, E^{\ast})$ offers a path to more directly separate sensitivities to intranuclear-cascade dynamics $-$ modeled in frameworks such as the \soft{Li\`{e}ge} intranuclear cascade model (INCL)~\cite{Boudard:2002yn,Boudard:2012wc}, \soft{DPMJET}~\cite{Ferrari:1995cq,Ferrari:1996fv}, or \soft{PEANUT}~\cite{Ferrari:2005zk,Ballarini:2024isa} $-$ from subsequent evaporation and fragmentation models (e.g., \soft{ABLA}~\cite{Kelic:2009yg}, \soft{FLUKA}~\cite{Ferrari:2005zk}, or related statistical frameworks), in reaction systems and beam-energy regimes that are comparatively underconstrained by existing data.\\
(iii) \emph{ultrarelativistic time dilation and boosted spectroscopy:} collider fragments emerge with Lorentz factors of order $\gamma \sim \mathcal{O}(10^2)$, extending the effective flight path of short-lived emitters. Consequently, even heavy $\alpha$ emitters or isomers with rest-frame lifetimes on the order of $\sim$ns can survive to far-forward instrumentation, and low-energy nuclear $\gamma$ rays are Doppler-boosted into the forward direction, enhancing detectability above typical machine-related soft-photon backgrounds.\\
(iv) \emph{absence of target energy loss and charge-state systematics:} unlike inverse-kinematics fixed-target measurements where energy loss, straggling, window corrections, and target-thickness uncertainties can smear the effective beam energy and complicate absolute cross sections~\cite{Taieb:2003jj,Bernas:2003hd}, collider operation eliminates ion energy loss in material targets. In addition, the fully stripped nature of the stored ion beams suppresses ambiguities associated with charge-state distributions~\cite{Haak:2023fbv,Ostroumov:2024kue}, reducing a class of systematic effects that can couple directly to rigidity-based fragment identification.
%------------------------------------------------------------

%------------------------------------------------------------
In this framework, the EIC does not replace the discovery reach or precision spectroscopy of existing rare-isotope facilities; instead, it complements them by introducing a differential, lepton-tagged view of excitation and de-excitation dynamics in the center-of-mass frame, with ultrarelativistic transport of the reaction products to far-forward detectors. This enables a coherent program in which isotope production and $\gamma$-spectroscopic signatures are correlated with well-defined initial-state kinematics and with controlled scans over nuclear species. Such correlations provide vital constraints on intranuclear transport and nuclear de-excitation models, which in turn inform targeted experimental campaigns at dedicated isotope facilities. The present work is motivated by this synergy, utilizing a realistic event-generator framework to explore how $e+A$ collisions at the EIC can produce highly neutron-rich pre-excited remnants, relate these pre-excited remnants to the largest experimentally accessible de-excited fragments, and leverage de-excitation $\gamma$ rays in a spectroscopic analysis to serve as unique fingerprints for specific isotopic identification. \\
%------------------------------------------------------------

%------------------------------------------------------------
%------------------------------------------------------------
%------------------------------------------------------------
\section*{Simulation Model} \label{sec:2}
%------------------------------------------------------------
This study utilizes version 1.03 of the Benchmark eA Generator for LEptoproduction (\beagle)~\cite{Chang:2022hkt}, a modular Monte Carlo (MC) event generator designed for high-energy lepton--nucleus scattering. \beagle integrates several established codes in a staged workflow to model the complete evolution of an interaction:\\

  \textsc{PYTHIA6}~\cite{Sjostrand:2006za}
  $\;\rightarrow\;$
  \textsc{PyQM}~\cite{Dupre:2011afa,Salgado:2003gb}
  $\;\rightarrow\;$
 \textsc{DPMJet}~\cite{Roesler:2000he}
  $\;\rightarrow\;$
  \textsc{FLUKA}~\cite{Ferrari:2005zk,Chang:2022hkt}.\\

These modules simulate the hard lepton--parton interaction and fragmentation~\cite{Sjostrand:2006za}, medium-induced partonic energy loss~\cite{Salgado:2003gb, Dupre:2011afa, Arleo:2018zjw}, intranuclear hadronic transport~\cite{Roesler:2000he}, and the statistical de-excitation of the exited nuclear remnant~\cite{Bohlen:2014buj, Ferrari:2005zk, Battistoni:2015epi}. Parton densities are provided via \soft{LHAPDF5}~\cite{Whalley:2005nh}. In this work, we used the model parameter set without additional tuning. For analysis, \beagle provides particle-level flags that identify the dominant production stage, enabling stage-aware selections tailored to the specific physics questions addressed herein.

The evolution of a typical electron--nucleus ($e+A$) collision in \beagle follows a chronological sequence that determines how the virtual photon ($\gamma^\ast$) deposits energy into the nucleus and how the resulting system de-excites. The process begins with the sampling of nuclear geometry and $\gamma^\ast$ kinematics. \beagle uses a Glauber-type~\cite{Miller:2007ri} initialization where the impact parameter, the position of the struck nucleon, and the spectator configuration are drawn from a nuclear density profile, such as a Woods--Saxon distribution. Simultaneously, the incoming lepton generates a virtual photon with a specific Bjorken-$x$ and four-momentum transfer $Q^2$, which mediates the primary interaction with a bound nucleon.

The primary lepton--nucleon scattering and initial fragmentation are then handled by \soft{PYTHIA6} using Deep-Inelastic-Scattering (DIS) processes and Lund string fragmentation~\cite{Sjostrand:2006za}. This stage produces an outgoing parton and a color string connecting it to the nuclear remnant. Subsequent string breaking and hadronization generate a set of final-state partons and pre-hadrons. Before complete hadronization, the \soft{PyQM} module accounts for medium-induced partonic energy loss. Using Salgado--Wiedemann quenching weights, the model redistributes parton energy to account for gluon radiation as partons traverse the nuclear medium~\cite{Dupre:2011afa, Arleo:2018zjw}. This quenching depends on the parton trajectory and the jet transport coefficient ($\hat{q}$), effectively depositing a portion of the photon's energy into the nuclear environment as soft gluons and recoiling nucleons.

As these objects evolve into hadrons, they propagate through the nucleus via \soft{DPMJet}~\cite{Roesler:2000he}. \beagle employs a formation-zone prescription where the formation time ($\tau_f$) determines when nascent hadrons begin to interact with spectator nucleons. In this work, $\tau_f$ is set to 5~fm/$c$, consistent with previous \beagle studies~\cite{Chang:2022hkt}. 
During the intranuclear cascade, secondary collisions drive the residual nucleus into a highly excited state. Following the intranuclear cascade, the system enters a pre-equilibrium state where energy is redistributed via exciton interactions. This transition leads to a statistical de-excitation phase, governed by the characteristic timescales illustrated in Fig.~\ref{fig:fig1}, once a thermalization condition is met and the system reaches an equilibrium state defined by $E^{\ast}$.

Finally, the excited remnant is passed to \soft{FLUKA} (version 2011.2b.5) for statistical de-excitation~\cite{Bohlen:2014buj, Ferrari:2005zk, Battistoni:2015epi}. \soft{FLUKA} simulates a sequence of emissions, including particle evaporation ($n, p, d, \alpha$), fission, or multifragmentation, alongside the emission of de-excitation $\gamma$ rays. This process continues until one or more cold nuclear fragments remain. We identify the largest of these as the ``biggest remnant'' to evaluate how final-state observables reflect the properties of the initial pre-excited system.

%------------------------------------------------------------
%------------------------------------------------------------
%--------------------------Results---------------------------
%------------------------------------------------------------
\section*{Results and Discussion}

Initially, we found it constructive to quantify the isotopic reach expected in electron–nucleus collisions at the future EIC using the \beagle model. Our first goal is to demonstrate that the virtual-photon flux in $e{+}A$ collisions can populate a broad range of excited nuclear systems, providing a natural framework for in-beam nuclear spectroscopy. 

%------------------------------------------------------------------------
\begin{figure}[htb]
\centering{
\includegraphics[width=1.0\linewidth]{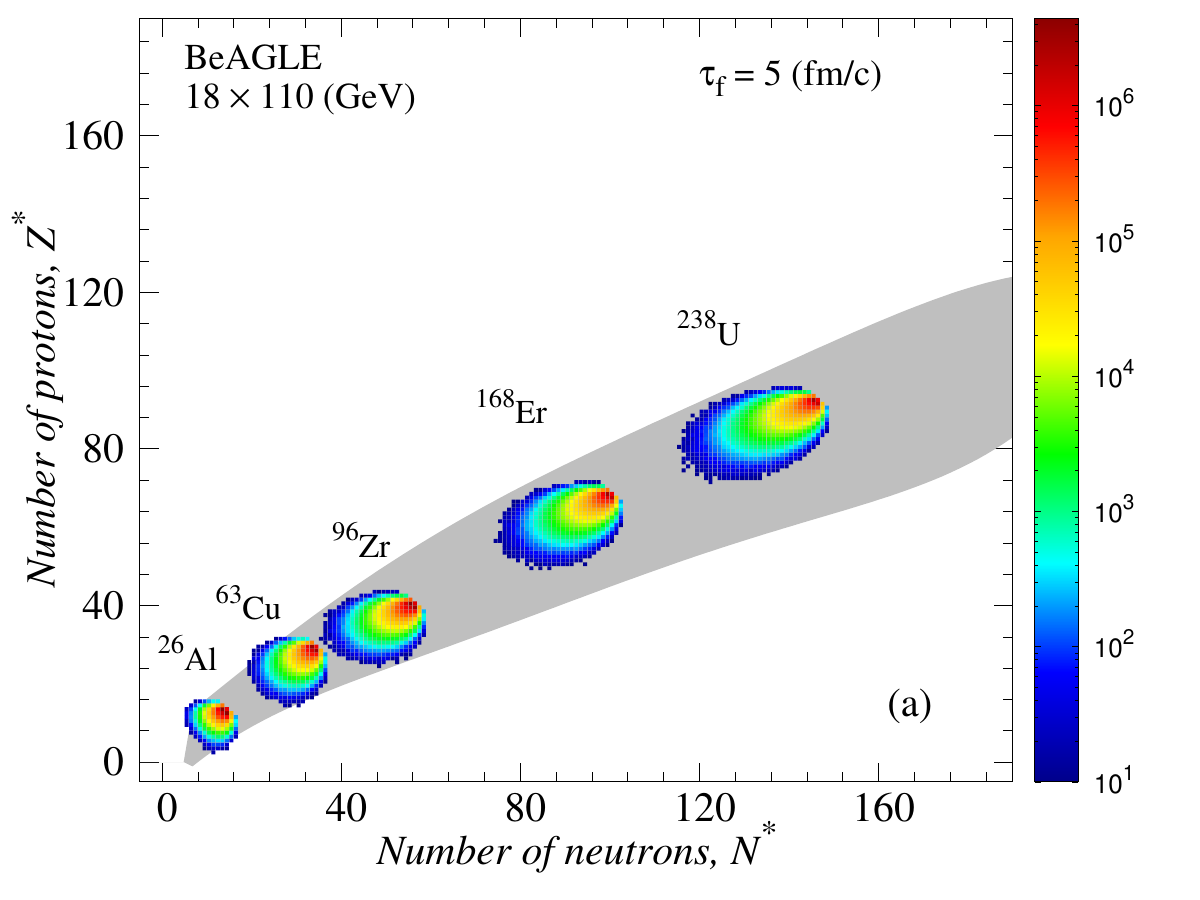}
\includegraphics[width=1.0\linewidth]{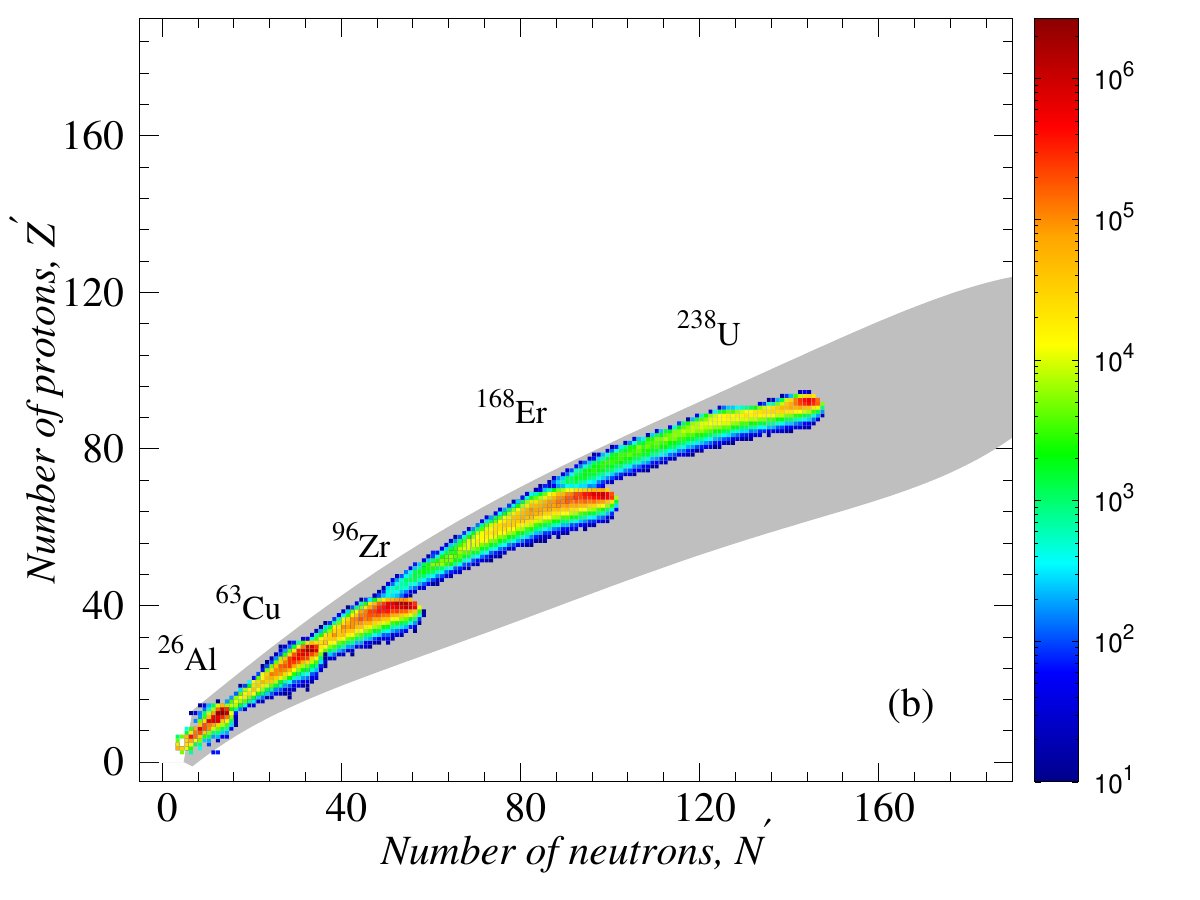}
\vskip -0.4cm
\caption{
Event-by-event correlations between the number of neutrons and protons for (a) excited nuclear remnant and (b) the largest residual nucleus, from Case-1, for $e{+}\text{A}$ at $18 \times 110$~GeV with $\tau_f = 5$~fm/$c$,  with no event cuts. The gray area illustrates the predicted range of bound nuclei extending out to the theoretical drip lines~\cite{Neufcourt:2020nme}.
%Phys.Rev.C 101 (2020) 4, 044307
\label{fig:fig2}
}
}
\vskip -0.1cm
\end{figure}
%------------------------------------------------------------------------

Figure~\ref{fig:fig2} provides a first validation of the central premise of this work: by varying the nuclear target (and, more generally, the beam conditions), $e+A$ collisions at EIC energies can populate a broad swath of the $(N, Z)$ landscape on an event-by-event basis. In Fig.~\ref{fig:fig2}(a), we show the event-level correlation between $N^{\ast}$ and $Z^{\ast}$ for the excited nuclear remnant, for several representative ion species at $\sqrt{s}=18\times110~\mathrm{GeV}$ (with $\tau_f=5~\mathrm{fm}/c$). Each ion species produces a distinct, finite-width distribution in the $(N^{\ast}, Z^{\ast})$ plane rather than a single point, reflecting fluctuations in nucleon removal, recoil, and energy deposition during the partonic and hadronic stages. Importantly, increasing the target size systematically shifts the populated region to larger $(N^{\ast}, Z^{\ast})$, effectively scanning different portions of the nuclear chart with the same collider kinematics. This behavior establishes the isotopic reach of the pre-equilibrium remnant: within a single beam configuration, the dynamics naturally generate an ensemble of excited systems with a spread in neutron and proton content around the original target.

The experimentally relevant question, however, is how this pre-equilibrium remnant is reflected in measurable final-state nuclei after statistical de-excitation. In this work, we emphasize Case-1 events, in which the remnant cools predominantly through evaporation/fragmentation, yielding a single dominant heavy residue accompanied by emitted hadrons. Figure~\ref{fig:fig2}(b) shows the corresponding $(N^{\prime}, Z^{\prime})$ distribution of the largest residual nucleus. Two qualitative features are evident. First, the post-de-excitation distributions remain broad and target-dependent, indicating that the isotopic diversity generated at the remnant stage is not washed out by the decay. Second, the mapping from $(N^{\ast}, Z^{\ast})$ to $(N^{\prime}, Z^{\prime})$ compresses the populated region into a narrower band, consistent with de-excitation preferentially removing excitation through light-particle emission while leaving a single heavy residue that carries most of the baryon number. In this sense, Case-1 events already provide a direct path to producing a family of heavy isotopes (the largest residues) whose yields and decay patterns can be studied further, while retaining sensitivity to the earlier dynamical stages that set $(A^{\ast}, Z^{\ast}, E^{\ast})$.

%------------------------------------------------------------------------
\begin{figure}[htb]
\centering{
\includegraphics[width=1.0\linewidth]{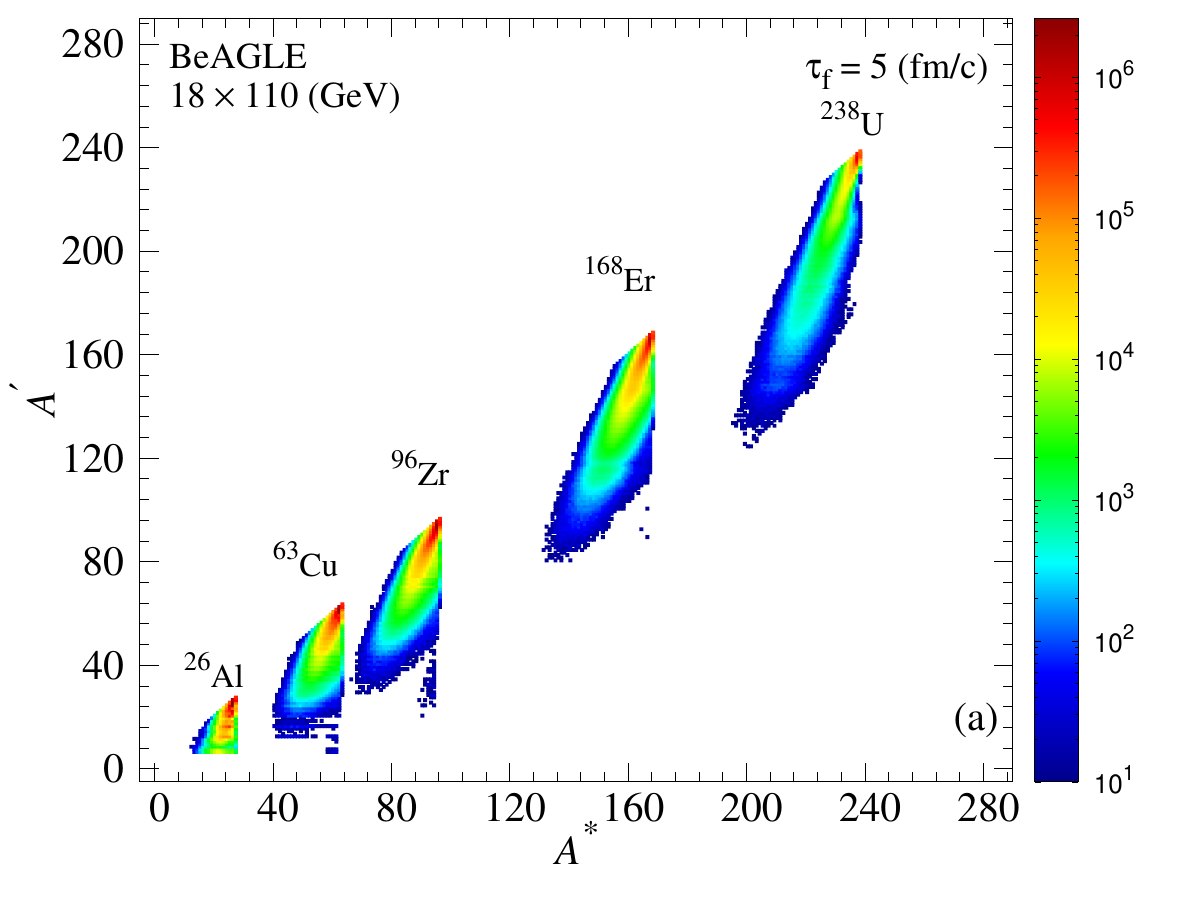}
\includegraphics[width=1.0\linewidth]{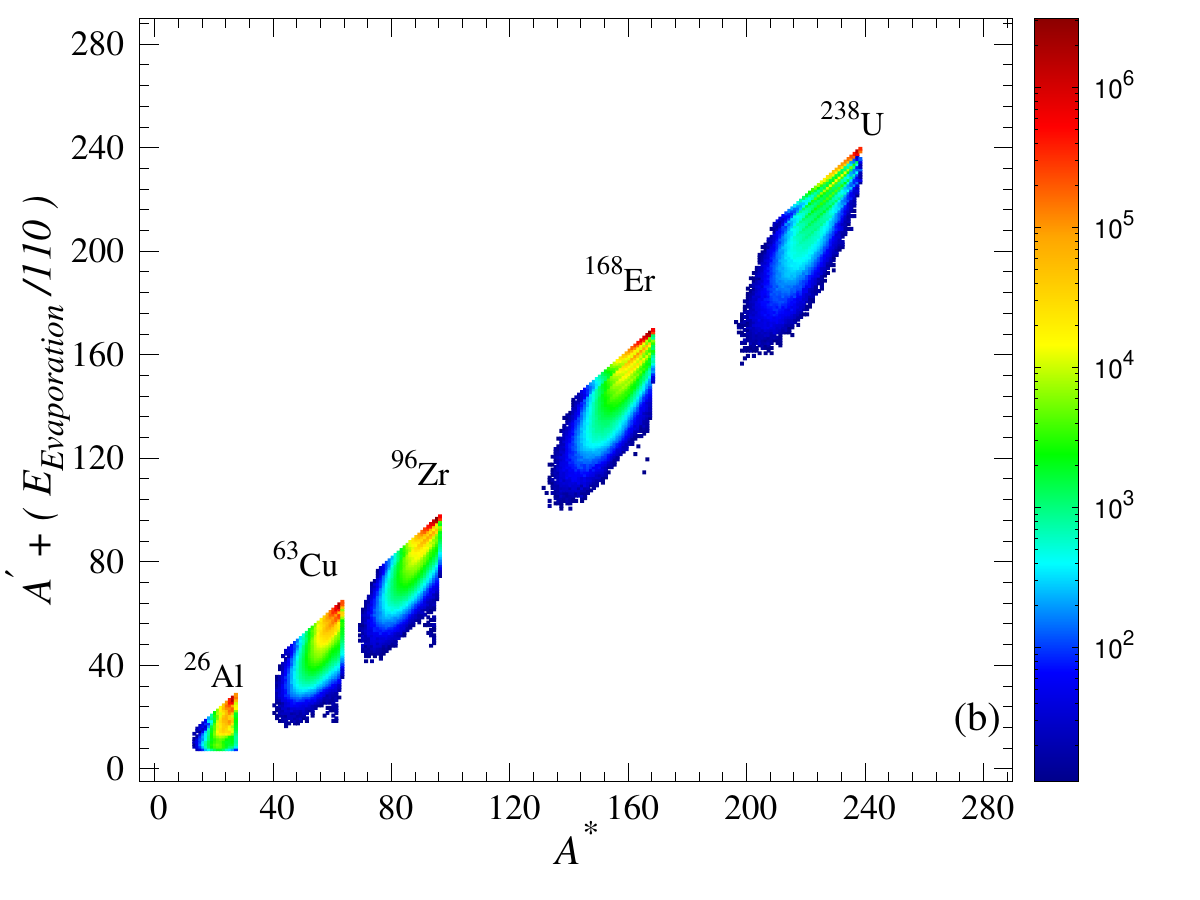}
\vskip -0.4cm
\caption{
Event-by-event correlations between the excited nuclear remnant ($A^{\ast}$) and the largest residual nucleus ($A^{\prime}$) panel (a) and the largest residual nucleus combined with the scaled evaporation neutron energy panel (b), from Case-1, for $e{+}\text{A}$ at $18 \times 110$~GeV with $\tau_f = 5$~fm/$c$. 
\label{fig:fig3}
}
}
\vskip -0.1cm
\end{figure}
%---------------------------------------------------

%---------------------------------------------------
While Fig.~\ref{fig:fig2} demonstrates isotopic reach, the excited remnant itself is not directly observable, motivating the need for practical, final-state proxies. Figure~\ref{fig:fig3}(a) shows that the mass number of the largest residual nucleus, $A^{\prime}$, is strongly correlated with the remnant mass number $A^{\ast}$ across all systems considered. The correlation is approximately linear, with event-by-event fluctuations around the dominant trend that encode variations in excitation and decay history. However, $A^{\prime}$ alone does not fully capture how much energy the remnant carried into de-excitation, since different events with similar $A^{\prime}$ can differ in the total evaporated energy and in the partition of excitation among emitted particles and internal degrees of freedom. This motivates augmenting the mass proxy by a measure of the evaporation energy, as shown in Fig.~\ref{fig:fig3}(b).

In Fig.~\ref{fig:fig3}(b), the $E_{\mathrm{evaporation}}$ is defined as the total energy of neutrons detected within the Zero Degree Calorimeter (ZDC) acceptance ($\eta > 6.0$)~\cite{AbdulKhalek:2021gbh, ePIC}.  With this $\eta$ acceptance, we can detect nearly all evaporation neutrons. By adding the scaled evaporation energy, $A^{\prime} + (E_{\mathrm{evaporation}}/110)$, we obtain a notably tighter and more nearly system-independent mapping to $A^{\ast}$. The scaling factor of 110~GeV corresponds to the nominal beam energy per nucleon carried by each forward-going neutron in the laboratory frame, allowing the total calorimetric energy to be converted directly into a neutron multiplicity. With this correction, the different ion species align along a common trend with reduced inter-system offset, and the residual deviations from linearity are small compared with the full dynamic range.  Practically, this suggests that a combined observable built from the largest fragment and an evaporation-energy estimator can serve as a calibrated handle on the unmeasured remnant mass, thereby enabling event-level constraints on the remnant properties most relevant to rare-isotope production and decay.
%---------------------------------------------------

%---------------------------------------------------
\begin{figure}[htb]
\centering{
\includegraphics[width=1.0\linewidth]{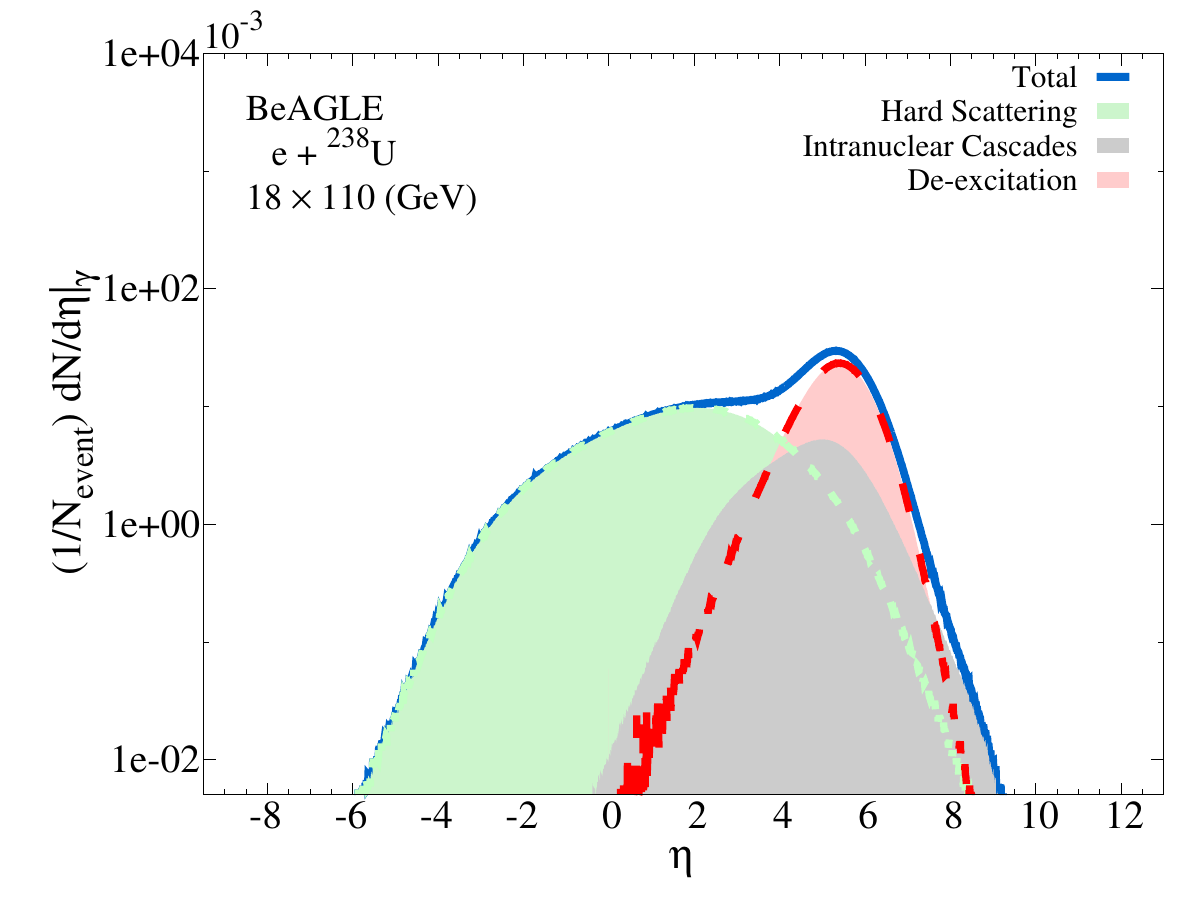}
\vskip -0.4cm
\caption{
Inclusive photon kinematics from \beagle at $18 \times 110$~GeV, shows the pseudorapidity distribution for $e{+}{}^{238}\text{U}$, decomposed into photons from hard scattering, intranuclear cascades, and nuclear de-excitation.
}\label{fig:fig4}
}
\end{figure}
%---------------------------------------------------

%---------------------------------------------------
Having established (i) a broad isotopic reach at the remnant stage and (ii) a strategy to relate that remnant to measurable final-state quantities, we turn to the spectroscopy-motivated question of whether de-excitation photons can be isolated experimentally from other photon sources in $e+A$ events. Figure~\ref{fig:fig4} shows the pseudorapidity distribution of photons in $e+{}^{238}\mathrm{U}$ at $18 \times 110~\mathrm{GeV}$, decomposed by production stage (hard scattering, intranuclear cascade, and nuclear de-excitation). The three components exhibit substantial overlap in rapidity, indicating that simple rapidity selections are unlikely to cleanly separate de-excitation photons from photons produced earlier in the reaction chain at these kinematics. This overlap motivates shifting the emphasis from longitudinal phase space to photon energy, where nuclear de-excitation via gamma-ray emission is expected to produce characteristic line-like structures.
%---------------------------------------------------

%---------------------------------------------------
\begin{figure}[htb]
\centering{
\includegraphics[width=1.0\linewidth]{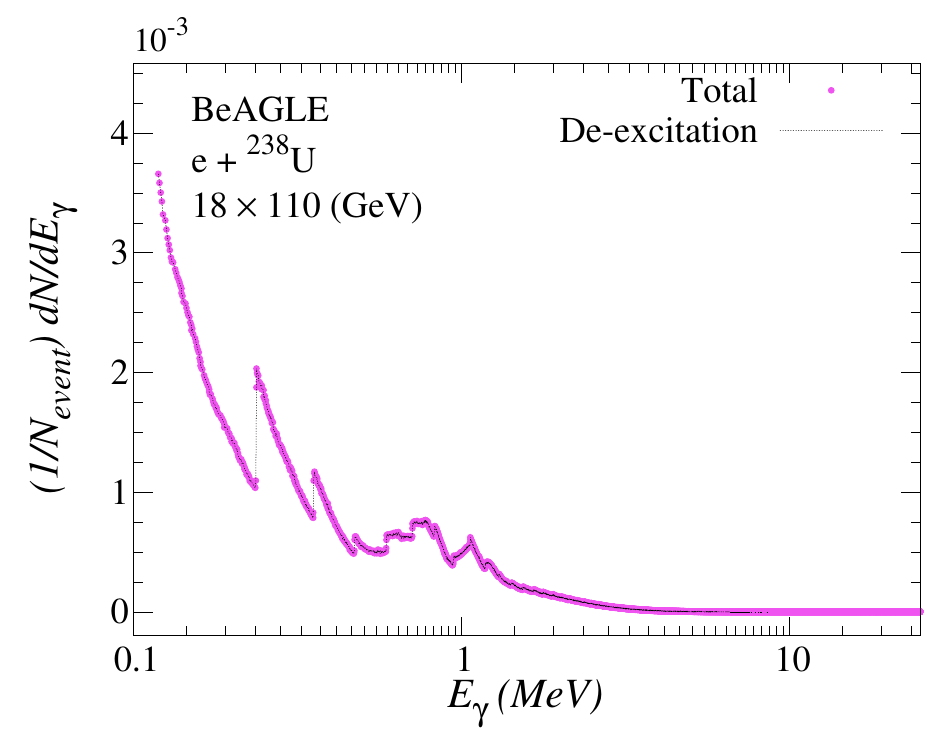}
\vskip -0.4cm
\caption{
Non-boosted photon–energy spectra for $e{+}{}^{238}\text{U}$, where $E_\gamma$ reflects the energy in the nucleus rest frame. The inclusive spectrum from all production stages (solid magenta) is compared with the specific contribution from nuclear de-excitation (dashed black). The de-excitation component dominates below $\sim 8$~MeV and exhibits discrete spectroscopic peaks.
}\label{fig:fig5}
}
\end{figure}
%---------------------------------------------------

%---------------------------------------------------
To isolate the intrinsic nuclear physics from the dominant Lorentz boost, Figure~\ref{fig:fig5} presents the de-excitation $\gamma$-energy spectrum transformed to the nuclear rest frame. In this frame, the low-energy $\gamma$-yield (below $\sim 8~\mathrm{MeV}$) is characterized by discrete peaks—the signature discrete transitions of nuclear de-excitation. Furthermore, the near-perfect correspondence between the total and de-excitation spectra (solid and dashed lines, respectively) confirms that de-excitation processes are the primary contributors to the $\gamma$-yield in this low-energy regime.
%---------------------------------------------------

%---------------------------------------------------
The discrete spectroscopic structures observed in Fig.~\ref{fig:fig5} reflect the integrated de-excitation profiles of the diverse isotopic ensemble shown in Fig.~\ref{fig:fig2}. In contrast, the inclusive photon spectrum includes additional, smoother contributions from hard and cascade processes that extend to higher energies. The separation in energy, rather than rapidity, thus provides the more promising lever arm for spectroscopy at the EIC: by combining event selections on the largest residual nucleus, and where possible additional constraints from evaporation observables that correlate with $A^{\ast}$, one can define ensembles of events that preferentially populate particular regions of $(A^{\ast}, Z^{\ast}, E^{\ast})$.
%---------------------------------------------------

%---------------------------------------------------
A logical next step for future work is to narrow the selections in $A^{\prime}$ and $Z^{\prime}$ to isolate isotope-specific features within the low-energy peak, thereby quantifying the extent to which the spectrum can be unfolded into contributions from specific excited nuclei and decay cascades. In practice, this unfolding requires rigorous control of both detector-level corrections and model dependence; while finite acceptance and resolution affect the measured neutron yield and fragment identification, any unmeasured light charged products make mapping the final state back to the prefragment configuration $(A^{\ast}, Z^{\ast}, E^{\ast})$ dependent on the assumed de-excitation treatment. Consequently, the resulting extraction of excited-nucleus populations will carry systematic uncertainties stemming from both experimental limitations and the chosen theoretical framework.
%---------------------------------------------------

%---------------------------------------------------
Finally, we note that the de-excitation of heavy remnants is not limited to Case-1 (one dominant residue plus hadrons). For sufficiently heavy and highly excited systems, binary breakup (Case-2), i.e., fission into two large fragments plus emitted hadrons, can compete and will populate a qualitatively different set of final nuclei, see Appendix.~\ref{app:a}. This channel both modifies the distribution of the largest fragment and opens additional opportunities for spectroscopy through the $\gamma$ cascades of the fission fragments and their decay correlations.

For each nuclear system presented in this study, we generated $10^7$ inelastic events. To place this statistical sample in the context of future EIC performance, we assume an instantaneous luminosity of $10^{33}$~$\text{cm}^{-2}\text{s}^{-1}$, corresponding to an inelastic $e+A$ collision rate of approximately $50$~\text{kHz} over the full $Q^{2} > 0$ range~\cite{AbdulKhalek:2021gbh}. Since each collision produces a single excited pre-fragment that subsequently decays through either the Case-1 or Case-2 channel, our simulated sample of 10 million events corresponds to roughly $200$~\text{s} of data collection. The relative roles of Case-1 and Case-2, and their impact on the observable fragment distributions and evaporation systematics, are further documented in the Appendix, providing complementary context for interpreting the correlations and photon spectra discussed above.
%---------------------------------------------------
%---------------------------------------------------

%---------------------------------------------------
%---------------------------------------------------
\section*{Summary and Conclusions}\label{sec:4}
%---------------------------------------------------
In this proof-of-principle study using the \beagle model, we demonstrated that $e{+}A$ collisions at the future EIC naturally populate a broad, target-dependent ensemble of pre-equilibrium remnant $(A^{\ast})$ and de-excited remnant $(A^{\prime})$. Because $(A^{\ast})$ is not directly observable, we established that combining the largest-fragment mass $A^{\prime}$ with a forward evaporation-energy estimator creates a robust, calibrated proxy for the remnant mass across different targets. Furthermore, we found that isolating de-excitation photons is experimentally feasible; while rapidity selections alone are insufficient due to background overlap, analyzing the photon-energy spectrum in the nuclear rest frame reveals discrete nuclear $\gamma$ cascades below $\sim 8$ MeV. Coupled with event selections based on fragment and evaporation observables, this energy-based approach provides a viable handle for nuclear spectroscopy at the EIC.

To develop these findings into an EIC-ready program, future work must systematically map the dependence of isotopic reach on collider kinematics, specifically scanning $(x, Q^{2})$ and beam energies, while rigorously quantifying uncertainties related to formation time, energy loss, and de-excitation models. Expanding the analysis to include binary breakup and fission topologies will be essential for heavy nuclei, alongside feasibility studies that incorporate full detector realism to translate generator-level correlations into robust experimental observables. In addition, comparing \beagle with \inclliege will establish a validated parameterization for \liseppcutesh simulations.

Ultimately, these efforts confirm that the EIC will function not just as a precision tool for nucleon structure, but as a novel source of excited nuclear systems. By systematically populating remote regions of the $(N, Z)$ plane and providing kinematically tagged reaction data, the EIC will serve as a powerful differential lens for exploring shell structure near drip lines, strongly complementing the discovery reach of dedicated Rare Isotope Beam facilities.

%------------------------------------------------------------------------
\section*{Acknowledgments}
%------------------------------------------------------------------------
The authors acknowledge using ChatGPT (OpenAI) for language revision and text improvement.
%For Oleg and Isaiah: 
This work was supported by the U.S. National Science Foundation under Grant No. 23-10078 (OBT and IR), PHY-2514907 (PNT), and by the U.S. Department of Energy under Grant No. DE-SC0024606 (MCH).
%------------------------------------------------------------------------

%\clearpage
\appendix

\subsection*{Appendix-A} \label{app:a}
\begin{figure}[htb]
\centering{
\includegraphics[width=1.0\linewidth]{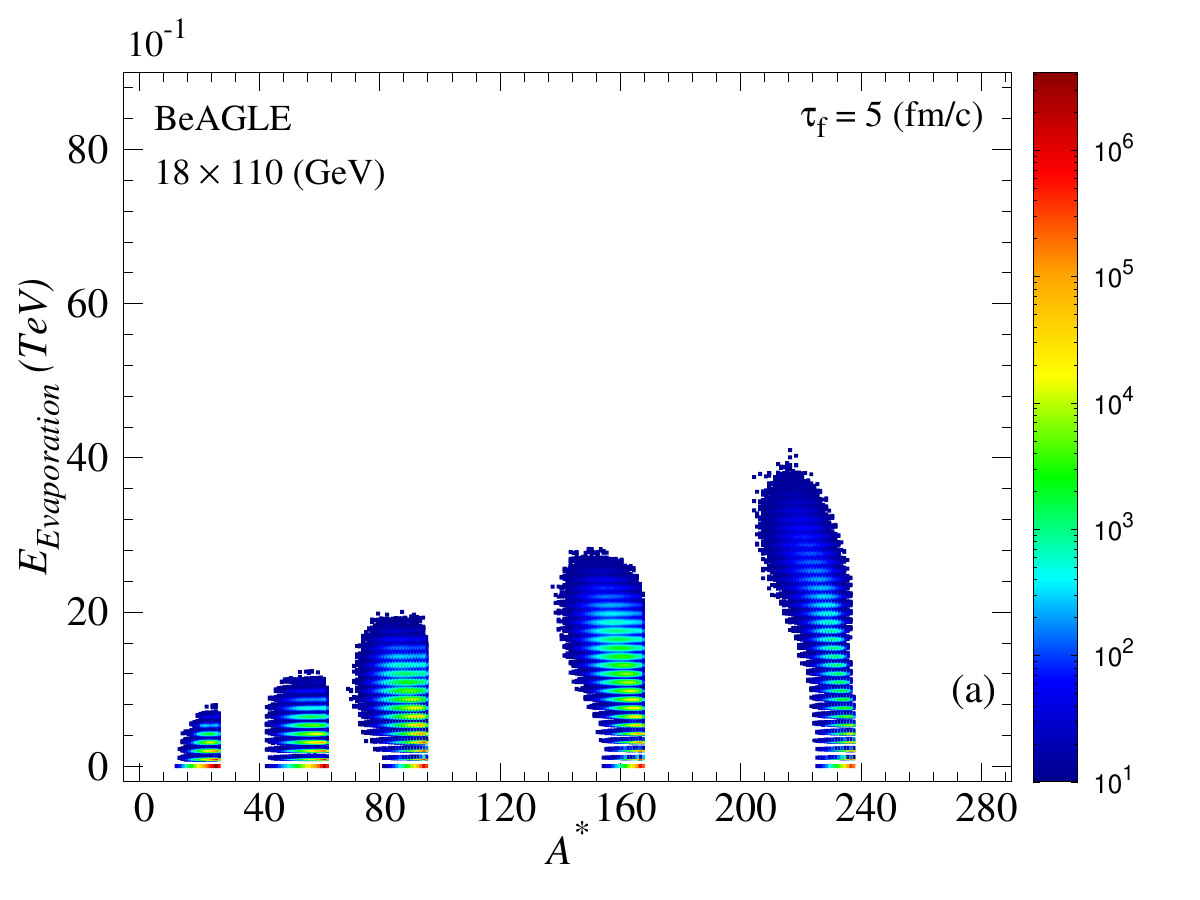}
\includegraphics[width=1.0\linewidth]{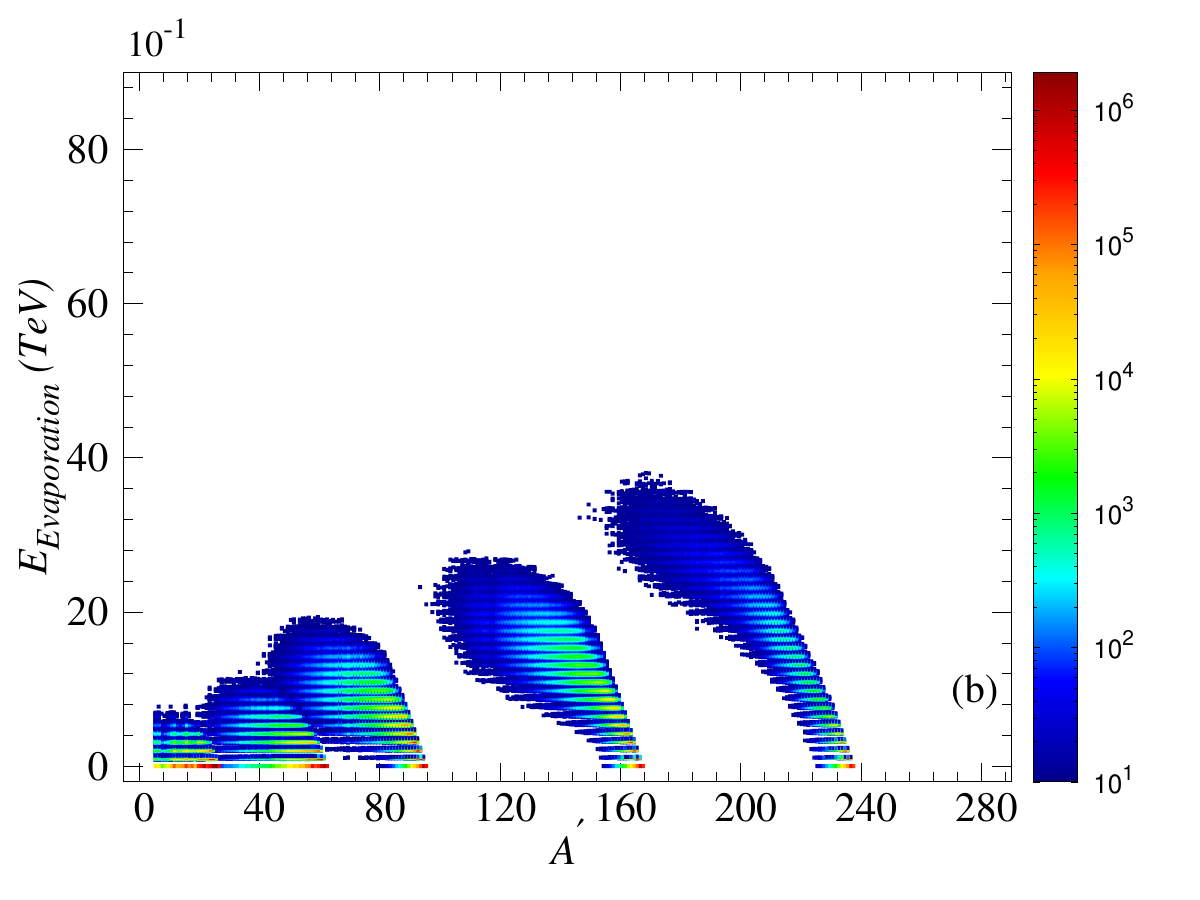}
\vskip -0.4cm
\caption{
Event-by-event correlations between the $A^{\ast}$ panel (a) and $A^{\prime}$ panel (b) with the evaporation neutron energy, from Case-1, for $e{+}\text{A}$ at $18 \times 110$~GeV with $\tau_f = 5$~fm/$c$.
\label{fig:figA1}
}
}
\vskip -0.1cm
\end{figure}

\begin{figure}[htb]
\centering{
\includegraphics[width=1.0\linewidth]{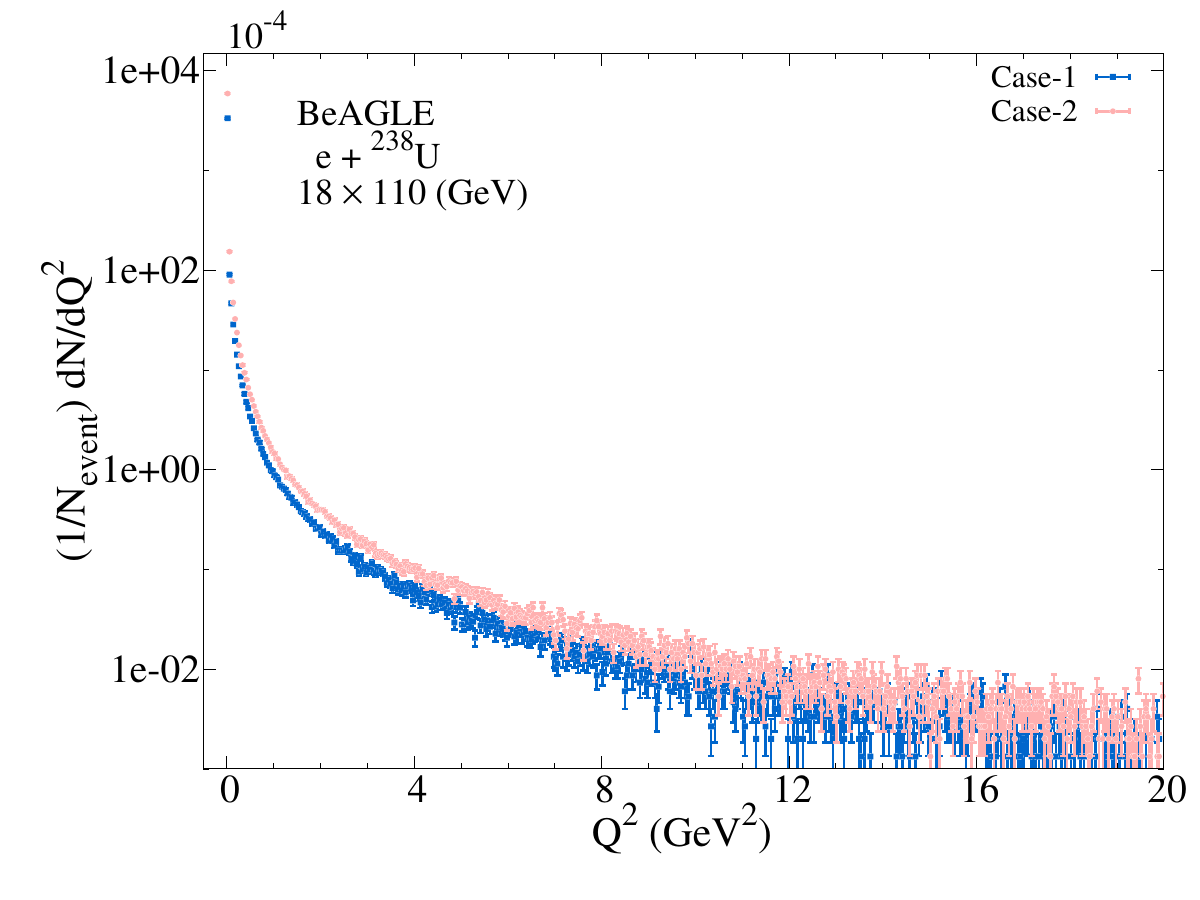}
\vskip -0.4cm
\caption{
The scaled $dN/dQ^{2}$ distribution for the two de-excitation mechanisms Case-1 and Case-2 for $e{+}^{238}\text{U}$ at $18 \times 110$~GeV with $\tau_f = 5$~fm/$c$.
\label{fig:figA2}
}
}
\vskip -0.1cm
\end{figure}

\begin{figure}[htb]
\centering{
\includegraphics[width=1.0\linewidth]{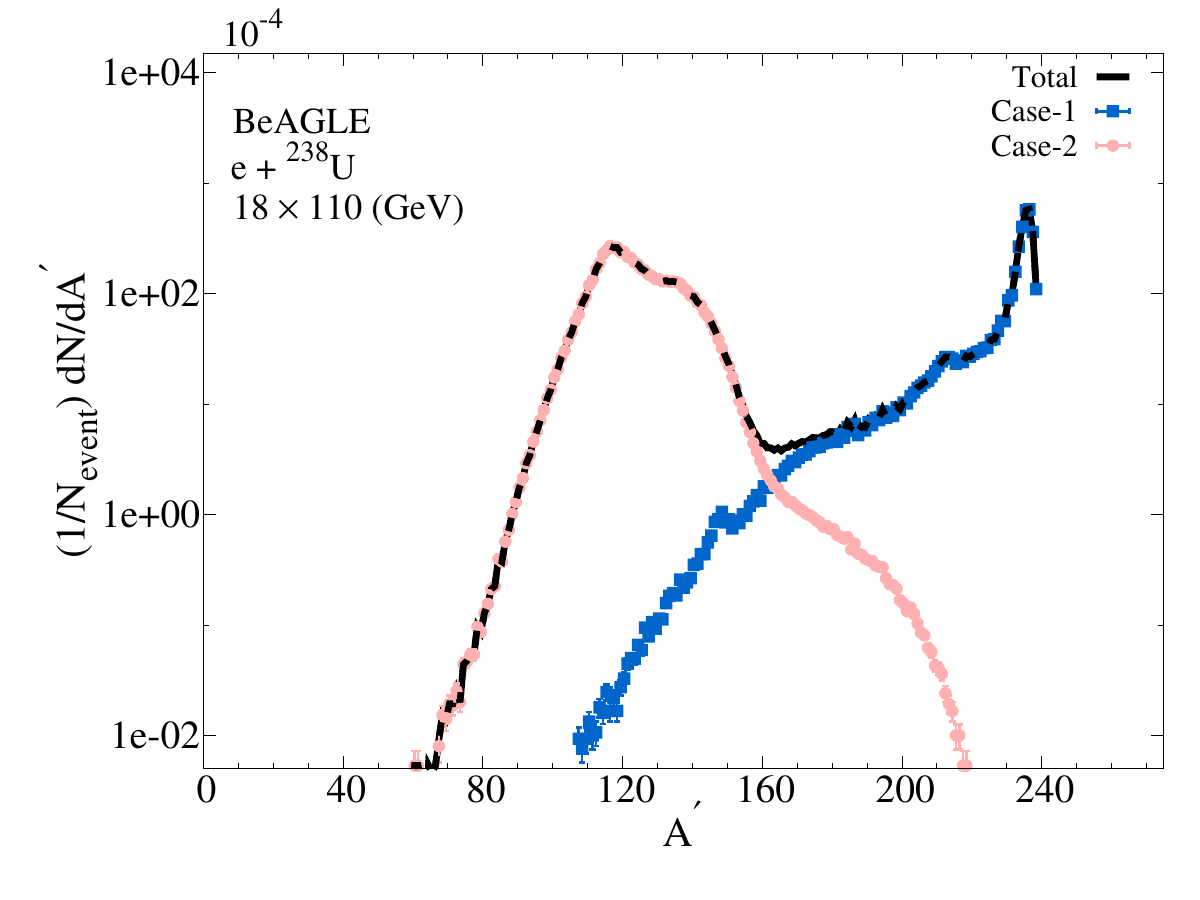}
\vskip -0.4cm
\caption{
The scaled $dN/dA^{\prime}$ distribution for the two de-excitation mechanisms Case-1 and Case-2 for $e{+}^{238}\text{U}$ at $18 \times 110$~GeV with $\tau_f = 5$~fm/$c$.
\label{fig:figA3}
}
}
\vskip -0.1cm
\end{figure}

This appendix shows (i) the event-by-event correlations between the $A^{\ast}$ and $A^{\prime}$ with the evaporation neutron energy Fig.~\ref{fig:figA1}, from Case-1, for $e{+}\text{A}$, (ii) the scaled $dN/dQ^{2}$ and $dN/dA^{\prime}$ distribution for the two de-excitation mechanisms Case-1 and Case-2 for $e{+}^{238}\text{U}$  Figs.~\ref{fig:figA2} and~\ref{fig:figA3}.

%------------------------------------------------------------------------

%------------------------------------------------------------------------

%------------------------------------------------------------------------
%\bibliographystyle{aipauth4-1}
\bibliography{ref2}

\begin{thebibliography}{56}%
\makeatletter
\providecommand \@ifxundefined [1]{%
 \@ifx{#1\undefined}
}%
\providecommand \@ifnum [1]{%
 \ifnum #1\expandafter \@firstoftwo
 \else \expandafter \@secondoftwo
 \fi
}%
\providecommand \@ifx [1]{%
 \ifx #1\expandafter \@firstoftwo
 \else \expandafter \@secondoftwo
 \fi
}%
\providecommand \natexlab [1]{#1}%
\providecommand \enquote  [1]{``#1''}%
\providecommand \bibnamefont  [1]{#1}%
\providecommand \bibfnamefont [1]{#1}%
\providecommand \citenamefont [1]{#1}%
\providecommand \href@noop [0]{\@secondoftwo}%
\providecommand \href [0]{\begingroup \@sanitize@url \@href}%
\providecommand \@href[1]{\@@startlink{#1}\@@href}%
\providecommand \@@href[1]{\endgroup#1\@@endlink}%
\providecommand \@sanitize@url [0]{\catcode `\\12\catcode `\$12\catcode
  `\&12\catcode `\#12\catcode `\^12\catcode `\_12\catcode `\%12\relax}%
\providecommand \@@startlink[1]{}%
\providecommand \@@endlink[0]{}%
\providecommand \url  [0]{\begingroup\@sanitize@url \@url }%
\providecommand \@url [1]{\endgroup\@href {#1}{\urlprefix }}%
\providecommand \urlprefix  [0]{URL }%
\providecommand \Eprint [0]{\href }%
\providecommand \doibase [0]{http://dx.doi.org/}%
\providecommand \selectlanguage [0]{\@gobble}%
\providecommand \bibinfo  [0]{\@secondoftwo}%
\providecommand \bibfield  [0]{\@secondoftwo}%
\providecommand \translation [1]{[#1]}%
\providecommand \BibitemOpen [0]{}%
\providecommand \bibitemStop [0]{}%
\providecommand \bibitemNoStop [0]{.\EOS\space}%
\providecommand \EOS [0]{\spacefactor3000\relax}%
\providecommand \BibitemShut  [1]{\csname bibitem#1\endcsname}%
\let\auto@bib@innerbib\@empty
%</preamble>
\bibitem [{\citenamefont {Abdul~Khalek}\ \emph {et~al.}(2022)\citenamefont
  {Abdul~Khalek} \emph {et~al.}}]{AbdulKhalek:2021gbh}%
  \BibitemOpen
  \bibfield  {author} {\bibinfo {author} {\bibfnamefont {R.}~\bibnamefont
  {Abdul~Khalek}} \emph {et~al.},\ }\href {\doibase
  10.1016/j.nuclphysa.2022.122447} {\bibfield  {journal} {\bibinfo  {journal}
  {Nucl. Phys. A}\ }\textbf {\bibinfo {volume} {1026}},\ \bibinfo {pages}
  {122447} (\bibinfo {year} {2022})},\ \Eprint
  {http://arxiv.org/abs/2103.05419} {arXiv:2103.05419 [physics.ins-det]}
  \BibitemShut {NoStop}%
\bibitem [{\citenamefont {Accardi}\ \emph {et~al.}(2016)\citenamefont {Accardi}
  \emph {et~al.}}]{Accardi:2012qut}%
  \BibitemOpen
  \bibfield  {author} {\bibinfo {author} {\bibfnamefont {A.}~\bibnamefont
  {Accardi}} \emph {et~al.},\ }\href {\doibase 10.1140/epja/i2016-16268-9}
  {\bibfield  {journal} {\bibinfo  {journal} {Eur. Phys. J. A}\ }\textbf
  {\bibinfo {volume} {52}},\ \bibinfo {pages} {268} (\bibinfo {year}
  {2016})}\BibitemShut {NoStop}%
\bibitem [{\citenamefont {Piller}\ \emph {et~al.}(1995)\citenamefont {Piller},
  \citenamefont {Ratzka},\ and\ \citenamefont {Weise}}]{Piller:1995kh}%
  \BibitemOpen
  \bibfield  {author} {\bibinfo {author} {\bibfnamefont {G.}~\bibnamefont
  {Piller}}, \bibinfo {author} {\bibfnamefont {W.}~\bibnamefont {Ratzka}}, \
  and\ \bibinfo {author} {\bibfnamefont {W.}~\bibnamefont {Weise}},\ }\href
  {\doibase 10.1007/BF01299761} {\bibfield  {journal} {\bibinfo  {journal} {Z.
  Phys. A}\ }\textbf {\bibinfo {volume} {352}},\ \bibinfo {pages} {427}
  (\bibinfo {year} {1995})},\ \Eprint {http://arxiv.org/abs/hep-ph/9504407}
  {arXiv:hep-ph/9504407} \BibitemShut {NoStop}%
\bibitem [{\citenamefont {Arneodo}(1994)}]{Arneodo:1994wf}%
  \BibitemOpen
  \bibfield  {author} {\bibinfo {author} {\bibfnamefont {M.}~\bibnamefont
  {Arneodo}},\ }\href {\doibase 10.1016/0370-1573(94)90048-5} {\bibfield
  {journal} {\bibinfo  {journal} {Phys. Rept.}\ }\textbf {\bibinfo {volume}
  {240}},\ \bibinfo {pages} {301} (\bibinfo {year} {1994})}\BibitemShut
  {NoStop}%
\bibitem [{\citenamefont {Cugnon}(1982)}]{Cugnon:1982qw}%
  \BibitemOpen
  \bibfield  {author} {\bibinfo {author} {\bibfnamefont {J.}~\bibnamefont
  {Cugnon}},\ }\href {\doibase 10.1016/0375-9474(82)90200-7} {\bibfield
  {journal} {\bibinfo  {journal} {Nucl. Phys. A}\ }\textbf {\bibinfo {volume}
  {387}},\ \bibinfo {pages} {191C} (\bibinfo {year} {1982})}\BibitemShut
  {NoStop}%
\bibitem [{\citenamefont {Weisskopf}(1937)}]{Weisskopf:1937zz}%
  \BibitemOpen
  \bibfield  {author} {\bibinfo {author} {\bibfnamefont {V.}~\bibnamefont
  {Weisskopf}},\ }\href {\doibase 10.1103/PhysRev.52.295} {\bibfield  {journal}
  {\bibinfo  {journal} {Phys. Rev.}\ }\textbf {\bibinfo {volume} {52}},\
  \bibinfo {pages} {295} (\bibinfo {year} {1937})}\BibitemShut {NoStop}%
\bibitem [{\citenamefont {Charity}(2010)}]{Charity:2010}%
  \BibitemOpen
  \bibfield  {author} {\bibinfo {author} {\bibfnamefont {R.~J.}\ \bibnamefont
  {Charity}},\ }\href {\doibase 10.1103/PhysRevC.82.014610} {\bibfield
  {journal} {\bibinfo  {journal} {Phys. Rev. C}\ }\textbf {\bibinfo {volume}
  {82}},\ \bibinfo {pages} {014610} (\bibinfo {year} {2010})}\BibitemShut
  {NoStop}%
\bibitem [{\citenamefont {Bondorf}\ \emph {et~al.}(1995)\citenamefont
  {Bondorf}, \citenamefont {Botvina}, \citenamefont {Iljinov}, \citenamefont
  {Mishustin},\ and\ \citenamefont {Sneppen}}]{Bondorf:1995ua}%
  \BibitemOpen
  \bibfield  {author} {\bibinfo {author} {\bibfnamefont {J.~P.}\ \bibnamefont
  {Bondorf}}, \bibinfo {author} {\bibfnamefont {A.~S.}\ \bibnamefont
  {Botvina}}, \bibinfo {author} {\bibfnamefont {A.~S.}\ \bibnamefont
  {Iljinov}}, \bibinfo {author} {\bibfnamefont {I.~N.}\ \bibnamefont
  {Mishustin}}, \ and\ \bibinfo {author} {\bibfnamefont {K.}~\bibnamefont
  {Sneppen}},\ }\href {\doibase 10.1016/0370-1573(94)00097-M} {\bibfield
  {journal} {\bibinfo  {journal} {Phys. Rept.}\ }\textbf {\bibinfo {volume}
  {257}},\ \bibinfo {pages} {133} (\bibinfo {year} {1995})}\BibitemShut
  {NoStop}%
\bibitem [{\citenamefont {Adderley}\ \emph {et~al.}(2024)\citenamefont
  {Adderley} \emph {et~al.}}]{Adderley:2024czm}%
  \BibitemOpen
  \bibfield  {author} {\bibinfo {author} {\bibfnamefont {P.~A.}\ \bibnamefont
  {Adderley}} \emph {et~al.},\ }\href {\doibase
  10.1103/PhysRevAccelBeams.27.084802} {\bibfield  {journal} {\bibinfo
  {journal} {Phys. Rev. Accel. Beams}\ }\textbf {\bibinfo {volume} {27}},\
  \bibinfo {pages} {084802} (\bibinfo {year} {2024})},\ \Eprint
  {http://arxiv.org/abs/2408.16880} {arXiv:2408.16880 [physics.acc-ph]}
  \BibitemShut {NoStop}%
\bibitem [{\citenamefont {Magdy}(2026)}]{Magdy:2025cse}%
  \BibitemOpen
  \bibfield  {author} {\bibinfo {author} {\bibfnamefont {N.}~\bibnamefont
  {Magdy}},\ }\href {\doibase 10.1103/kqf6-xczd} {\bibfield  {journal}
  {\bibinfo  {journal} {Phys. Rev. C}\ }\textbf {\bibinfo {volume} {113}},\
  \bibinfo {pages} {L011901} (\bibinfo {year} {2026})},\ \Eprint
  {http://arxiv.org/abs/2506.07426} {arXiv:2506.07426 [hep-ph]} \BibitemShut
  {NoStop}%
\bibitem [{\citenamefont {Crawford}\ \emph {et~al.}(2023)\citenamefont
  {Crawford}, \citenamefont {Fossez}, \citenamefont {K{\"o}nig},\ and\
  \citenamefont {Spyrou}}]{Crawford:2023txq}%
  \BibitemOpen
  \bibfield  {author} {\bibinfo {author} {\bibfnamefont {H.~L.}\ \bibnamefont
  {Crawford}}, \bibinfo {author} {\bibfnamefont {K.}~\bibnamefont {Fossez}},
  \bibinfo {author} {\bibfnamefont {S.}~\bibnamefont {K{\"o}nig}}, \ and\
  \bibinfo {author} {\bibfnamefont {A.}~\bibnamefont {Spyrou}},\ }\href
  {\doibase 10.1146/annurev-nucl-121423-091501} {\  (\bibinfo {year} {2023}),\
  10.1146/annurev-nucl-121423-091501},\ \Eprint
  {http://arxiv.org/abs/2312.09129} {arXiv:2312.09129 [nucl-ex]} \BibitemShut
  {NoStop}%
\bibitem [{\citenamefont {Horowitz}\ \emph {et~al.}(2019)\citenamefont
  {Horowitz} \emph {et~al.}}]{Horowitz:2018ndv}%
  \BibitemOpen
  \bibfield  {author} {\bibinfo {author} {\bibfnamefont {C.~J.}\ \bibnamefont
  {Horowitz}} \emph {et~al.},\ }\href {\doibase 10.1088/1361-6471/ab0849}
  {\bibfield  {journal} {\bibinfo  {journal} {J. Phys. G}\ }\textbf {\bibinfo
  {volume} {46}},\ \bibinfo {pages} {083001} (\bibinfo {year} {2019})},\
  \Eprint {http://arxiv.org/abs/1805.04637} {arXiv:1805.04637 [astro-ph.SR]}
  \BibitemShut {NoStop}%
\bibitem [{\citenamefont {{National Nuclear Data Center}}()}]{NNDC_NuDat}%
  \BibitemOpen
  \bibfield  {author} {\bibinfo {author} {\bibnamefont {{National Nuclear Data
  Center}}},\ }\href@noop {} {\enquote {\bibinfo {title} {{NuDat}~3: Nuclear
  structure and decay data},}\ }\bibinfo {howpublished}
  {\url{https://www.nndc.bnl.gov/nudat3/}},\ \bibinfo {note} {accessed 26 Oct
  2025}\BibitemShut {NoStop}%
\bibitem [{\citenamefont {Neufcourt}\ \emph {et~al.}(2020)\citenamefont
  {Neufcourt}, \citenamefont {Cao}, \citenamefont {Giuliani}, \citenamefont
  {Nazarewicz}, \citenamefont {Olsen},\ and\ \citenamefont
  {Tarasov}}]{Neufcourt:2020nme}%
  \BibitemOpen
  \bibfield  {author} {\bibinfo {author} {\bibfnamefont {L.}~\bibnamefont
  {Neufcourt}}, \bibinfo {author} {\bibfnamefont {Y.}~\bibnamefont {Cao}},
  \bibinfo {author} {\bibfnamefont {S.~A.}\ \bibnamefont {Giuliani}}, \bibinfo
  {author} {\bibfnamefont {W.}~\bibnamefont {Nazarewicz}}, \bibinfo {author}
  {\bibfnamefont {E.}~\bibnamefont {Olsen}}, \ and\ \bibinfo {author}
  {\bibfnamefont {O.~B.}\ \bibnamefont {Tarasov}},\ }\href {\doibase
  10.1103/PhysRevC.101.044307} {\bibfield  {journal} {\bibinfo  {journal}
  {Phys. Rev. C}\ }\textbf {\bibinfo {volume} {101}},\ \bibinfo {pages}
  {044307} (\bibinfo {year} {2020})},\ \Eprint
  {http://arxiv.org/abs/2001.05924} {arXiv:2001.05924 [nucl-th]} \BibitemShut
  {NoStop}%
\bibitem [{\citenamefont {{International Atomic Energy
  Agency}}()}]{IAEA_LiveChart}%
  \BibitemOpen
  \bibfield  {author} {\bibinfo {author} {\bibnamefont {{International Atomic
  Energy Agency}}},\ }\href@noop {} {\enquote {\bibinfo {title} {Live chart of
  nuclides},}\ }\bibinfo {howpublished}
  {\url{https://www-nds.iaea.org/relnsd/vcharthtml/VChartHTML.html}},\ \bibinfo
  {note} {accessed 26 Oct 2025}\BibitemShut {NoStop}%
\bibitem [{\citenamefont {{Facility for Rare Isotope Beams}}()}]{FRIB}%
  \BibitemOpen
  \bibfield  {author} {\bibinfo {author} {\bibnamefont {{Facility for Rare
  Isotope Beams}}},\ }\href@noop {} {\enquote {\bibinfo {title} {Facility for
  rare isotope beams (frib) at michigan state university},}\ }\bibinfo
  {howpublished} {\url{https://frib.msu.edu/}},\ \bibinfo {note} {accessed 26
  Oct 2025}\BibitemShut {NoStop}%
\bibitem [{\citenamefont {{U.S. Department of Energy Office of Science, Nuclear
  Physics}}(2025)}]{DOEFRIB}%
  \BibitemOpen
  \bibfield  {author} {\bibinfo {author} {\bibnamefont {{U.S. Department of
  Energy Office of Science, Nuclear Physics}}},\ }\href@noop {} {\enquote
  {\bibinfo {title} {Facility for rare isotope beams (frib) --- user facility
  overview},}\ }\bibinfo {howpublished}
  {\url{https://science.osti.gov/np/Facilities/User-Facilities/FRIB}} (\bibinfo
  {year} {2025}),\ \bibinfo {note} {accessed 26 Oct 2025}\BibitemShut {NoStop}%
\bibitem [{\citenamefont {{U.S. Department of Energy, Nuclear
  Physics}}(2023)}]{FRIB_OneYear}%
  \BibitemOpen
  \bibfield  {author} {\bibinfo {author} {\bibnamefont {{U.S. Department of
  Energy, Nuclear Physics}}},\ }\href@noop {} {\enquote {\bibinfo {title} {The
  facility for rare isotope beams after one year of operation},}\ }\bibinfo
  {howpublished}
  {\url{https://www.energy.gov/science/np/articles/facility-rare-isotope-beams-after-one-year-operation}}
  (\bibinfo {year} {2023}),\ \bibinfo {note} {accessed 26 Oct 2025}\BibitemShut
  {NoStop}%
\bibitem [{\citenamefont {{RIKEN Nishina Center}}()}]{RIBF}%
  \BibitemOpen
  \bibfield  {author} {\bibinfo {author} {\bibnamefont {{RIKEN Nishina
  Center}}},\ }\href@noop {} {\enquote {\bibinfo {title} {What is the
  {RIBF}?}}\ }\bibinfo {howpublished}
  {\url{https://www.nishina.riken.jp/facility/RIBFabout_e.html}},\ \bibinfo
  {note} {accessed 26 Oct 2025}\BibitemShut {NoStop}%
\bibitem [{\citenamefont {{CERN ISOLDE}}(2025)}]{ISOLDE}%
  \BibitemOpen
  \bibfield  {author} {\bibinfo {author} {\bibnamefont {{CERN ISOLDE}}},\
  }\href@noop {} {\enquote {\bibinfo {title} {The {ISOLDE} radioactive ion beam
  facility},}\ }\bibinfo {howpublished}
  {\url{https://isolde.cern/isolde-radioactive-ion-beam-facility}} (\bibinfo
  {year} {2025}),\ \bibinfo {note} {accessed 26 Oct 2025}\BibitemShut {NoStop}%
\bibitem [{\citenamefont {{CERN ISOLDE}}()}]{ISOLDEsetups}%
  \BibitemOpen
  \bibfield  {author} {\bibinfo {author} {\bibnamefont {{CERN ISOLDE}}},\
  }\href@noop {} {\enquote {\bibinfo {title} {Experimental setups at
  {ISOLDE}},}\ }\bibinfo {howpublished}
  {\url{https://isolde.cern/experimental-setups}},\ \bibinfo {note} {accessed
  26 Oct 2025}\BibitemShut {NoStop}%
\bibitem [{\citenamefont {{GANIL--SPIRAL2}}()}]{GANILSPIRAL2}%
  \BibitemOpen
  \bibfield  {author} {\bibinfo {author} {\bibnamefont {{GANIL--SPIRAL2}}},\
  }\href@noop {} {\enquote {\bibinfo {title} {Accelerators at
  {GANIL--SPIRAL2}},}\ }\bibinfo {howpublished}
  {\url{https://www.ganil-spiral2.eu/scientists/ganil-spiral-2-facilities/accelerators/}},\
  \bibinfo {note} {accessed 26 Oct 2025}\BibitemShut {NoStop}%
\bibitem [{\citenamefont {{FAIR/GSI}}()}]{FAIR}%
  \BibitemOpen
  \bibfield  {author} {\bibinfo {author} {\bibnamefont {{FAIR/GSI}}},\
  }\href@noop {} {\enquote {\bibinfo {title} {Superconducting fragment
  separator (super-{FRS})},}\ }\bibinfo {howpublished}
  {\url{https://fair-center.de/user/experiments/nustar/super-frs}},\ \bibinfo
  {note} {accessed 26 Oct 2025}\BibitemShut {NoStop}%
\bibitem [{\citenamefont {{FAIR Center}}()}]{FAIR_NUSTAR}%
  \BibitemOpen
  \bibfield  {author} {\bibinfo {author} {\bibnamefont {{FAIR Center}}},\
  }\href@noop {} {\enquote {\bibinfo {title} {{NUSTAR}: Nuclear structure,
  astrophysics and reactions},}\ }\bibinfo {howpublished}
  {\url{https://fair-center.de/user/experiments/nustar}},\ \bibinfo {note}
  {accessed 26 Oct 2025}\BibitemShut {NoStop}%
\bibitem [{\citenamefont {{TRIUMF}}({\natexlab{a}})}]{TRIUMF_ARIEL}%
  \BibitemOpen
  \bibfield  {author} {\bibinfo {author} {\bibnamefont {{TRIUMF}}},\
  }\href@noop {} {\enquote {\bibinfo {title} {{ARIEL} --- advanced rare isotope
  laboratory},}\ }\bibinfo {howpublished}
  {\url{https://www.triumf.ca/facilities-experiments/ariel/}}
  ({\natexlab{a}}),\ \bibinfo {note} {accessed 26 Oct 2025}\BibitemShut
  {NoStop}%
\bibitem [{\citenamefont {{TRIUMF}}({\natexlab{b}})}]{TRIUMF_TIGRESS}%
  \BibitemOpen
  \bibfield  {author} {\bibinfo {author} {\bibnamefont {{TRIUMF}}},\
  }\href@noop {} {\enquote {\bibinfo {title} {{TIGRESS}: {TRIUMF}-{ISAC}
  gamma-ray escape suppressed spectrometer},}\ }\bibinfo {howpublished}
  {\url{https://fiveyearplan.triumf.ca/teams-tools/tigress-triumf-isac-gamma-ray-suppressed-spectrometer/index.html}}
  ({\natexlab{b}}),\ \bibinfo {note} {accessed 26 Oct 2025}\BibitemShut
  {NoStop}%
\bibitem [{\citenamefont {{Argonne National
  Laboratory}}({\natexlab{a}})}]{ATLAS_CARIBU}%
  \BibitemOpen
  \bibfield  {author} {\bibinfo {author} {\bibnamefont {{Argonne National
  Laboratory}}},\ }\href@noop {} {\enquote {\bibinfo {title} {Californium rare
  isotope breeder upgrade ({CARIBU})},}\ }\bibinfo {howpublished}
  {\url{https://www.anl.gov/phy/californium-rare-isotope-breeder-upgrade-caribu}}
  ({\natexlab{a}}),\ \bibinfo {note} {accessed 26 Oct 2025}\BibitemShut
  {NoStop}%
\bibitem [{\citenamefont {{Argonne National
  Laboratory}}({\natexlab{b}})}]{nuCARIBU}%
  \BibitemOpen
  \bibfield  {author} {\bibinfo {author} {\bibnamefont {{Argonne National
  Laboratory}}},\ }\href@noop {} {\enquote {\bibinfo {title} {{nuCARIBU}
  beams},}\ }\bibinfo {howpublished}
  {\url{https://www.anl.gov/atlas/nucaribu-beams}} ({\natexlab{b}}),\ \bibinfo
  {note} {accessed 26 Oct 2025}\BibitemShut {NoStop}%
\bibitem [{\citenamefont {{University of Jyväskylä}}()}]{JYFL_IGISOL}%
  \BibitemOpen
  \bibfield  {author} {\bibinfo {author} {\bibnamefont {{University of
  Jyväskylä}}},\ }\href@noop {} {\enquote {\bibinfo {title} {Exotic nuclei
  and beams ({IGISOL})},}\ }\bibinfo {howpublished}
  {\url{https://www.jyu.fi/en/research-groups/exotic-nuclei-and-beams-igisol}},\
  \bibinfo {note} {accessed 26 Oct 2025}\BibitemShut {NoStop}%
\bibitem [{\citenamefont {{ALICE Collaboration}}(2012)}]{UPC_Pb208}%
  \BibitemOpen
  \bibfield  {author} {\bibinfo {author} {\bibnamefont {{ALICE
  Collaboration}}},\ }\href {\doibase 10.1103/PhysRevLett.109.252302}
  {\bibfield  {journal} {\bibinfo  {journal} {Phys. Rev. Lett.}\ }\textbf
  {\bibinfo {volume} {109}},\ \bibinfo {pages} {252302} (\bibinfo {year}
  {2012})}\BibitemShut {NoStop}%
\bibitem [{\citenamefont {{Brookhaven National Laboratory}}()}]{BNL_RHICvsEIC}%
  \BibitemOpen
  \bibfield  {author} {\bibinfo {author} {\bibnamefont {{Brookhaven National
  Laboratory}}},\ }\href@noop {} {\enquote {\bibinfo {title} {{RHIC} and the
  {EIC}},}\ }\bibinfo {howpublished}
  {\url{https://www.bnl.gov/eic/rhic-eic.php}},\ \bibinfo {note} {accessed 26
  Oct 2025}\BibitemShut {NoStop}%
\bibitem [{\citenamefont {Alarcon}\ \emph {et~al.}(2022)\citenamefont {Alarcon}
  \emph {et~al.}}]{CORE:2022rso}%
  \BibitemOpen
  \bibfield  {author} {\bibinfo {author} {\bibfnamefont {R.}~\bibnamefont
  {Alarcon}} \emph {et~al.} (\bibinfo {collaboration} {CORE}),\ }\href@noop {}
  {\  (\bibinfo {year} {2022})},\ \Eprint {http://arxiv.org/abs/2209.00496}
  {arXiv:2209.00496 [physics.ins-det]} \BibitemShut {NoStop}%
\bibitem [{\citenamefont {Bertulani}\ \emph {et~al.}(2025)\citenamefont
  {Bertulani}, \citenamefont {Kucuk},\ and\ \citenamefont
  {Navarra}}]{Bertulani:2024mqe}%
  \BibitemOpen
  \bibfield  {author} {\bibinfo {author} {\bibfnamefont {C.~A.}\ \bibnamefont
  {Bertulani}}, \bibinfo {author} {\bibfnamefont {Y.}~\bibnamefont {Kucuk}}, \
  and\ \bibinfo {author} {\bibfnamefont {F.~S.}\ \bibnamefont {Navarra}},\
  }\href {\doibase 10.1016/j.nuclphysa.2025.123093} {\bibfield  {journal}
  {\bibinfo  {journal} {Nucl. Phys. A}\ }\textbf {\bibinfo {volume} {1059}},\
  \bibinfo {pages} {123093} (\bibinfo {year} {2025})},\ \Eprint
  {http://arxiv.org/abs/2408.10157} {arXiv:2408.10157 [nucl-th]} \BibitemShut
  {NoStop}%
\bibitem [{\citenamefont {Kim}\ \emph {et~al.}(2026)\citenamefont {Kim} \emph
  {et~al.}}]{Kim:2026ytt}%
  \BibitemOpen
  \bibfield  {author} {\bibinfo {author} {\bibfnamefont {J.}~\bibnamefont
  {Kim}} \emph {et~al.},\ }\href@noop {} {\  (\bibinfo {year} {2026})},\
  \Eprint {http://arxiv.org/abs/2602.04636} {arXiv:2602.04636 [nucl-ex]}
  \BibitemShut {NoStop}%
\bibitem [{\citenamefont {Taieb}\ \emph {et~al.}(2003)\citenamefont {Taieb}
  \emph {et~al.}}]{Taieb:2003jj}%
  \BibitemOpen
  \bibfield  {author} {\bibinfo {author} {\bibfnamefont {J.}~\bibnamefont
  {Taieb}} \emph {et~al.},\ }\href {\doibase 10.1016/S0375-9474(03)01517-3}
  {\bibfield  {journal} {\bibinfo  {journal} {Nucl. Phys. A}\ }\textbf
  {\bibinfo {volume} {724}},\ \bibinfo {pages} {413} (\bibinfo {year}
  {2003})},\ \Eprint {http://arxiv.org/abs/nucl-ex/0302026}
  {arXiv:nucl-ex/0302026} \BibitemShut {NoStop}%
\bibitem [{\citenamefont {Bernas}\ \emph {et~al.}(2003)\citenamefont {Bernas}
  \emph {et~al.}}]{Bernas:2003hd}%
  \BibitemOpen
  \bibfield  {author} {\bibinfo {author} {\bibfnamefont {M.}~\bibnamefont
  {Bernas}} \emph {et~al.},\ }\href {\doibase 10.1016/S0375-9474(03)01576-8}
  {\bibfield  {journal} {\bibinfo  {journal} {Nucl. Phys. A}\ }\textbf
  {\bibinfo {volume} {725}},\ \bibinfo {pages} {213} (\bibinfo {year}
  {2003})},\ \Eprint {http://arxiv.org/abs/nucl-ex/0304003}
  {arXiv:nucl-ex/0304003} \BibitemShut {NoStop}%
\bibitem [{\citenamefont {Boudard}\ \emph {et~al.}(2002)\citenamefont
  {Boudard}, \citenamefont {Cugnon}, \citenamefont {Leray},\ and\ \citenamefont
  {Volant}}]{Boudard:2002yn}%
  \BibitemOpen
  \bibfield  {author} {\bibinfo {author} {\bibfnamefont {A.}~\bibnamefont
  {Boudard}}, \bibinfo {author} {\bibfnamefont {J.}~\bibnamefont {Cugnon}},
  \bibinfo {author} {\bibfnamefont {S.}~\bibnamefont {Leray}}, \ and\ \bibinfo
  {author} {\bibfnamefont {C.}~\bibnamefont {Volant}},\ }\href {\doibase
  10.1103/PhysRevC.66.044615} {\bibfield  {journal} {\bibinfo  {journal} {Phys.
  Rev. C}\ }\textbf {\bibinfo {volume} {66}},\ \bibinfo {pages} {044615}
  (\bibinfo {year} {2002})}\BibitemShut {NoStop}%
\bibitem [{\citenamefont {Boudard}\ \emph {et~al.}(2013)\citenamefont
  {Boudard}, \citenamefont {Cugnon}, \citenamefont {David}, \citenamefont
  {Leray},\ and\ \citenamefont {Mancusi}}]{Boudard:2012wc}%
  \BibitemOpen
  \bibfield  {author} {\bibinfo {author} {\bibfnamefont {A.}~\bibnamefont
  {Boudard}}, \bibinfo {author} {\bibfnamefont {J.}~\bibnamefont {Cugnon}},
  \bibinfo {author} {\bibfnamefont {J.-C.}\ \bibnamefont {David}}, \bibinfo
  {author} {\bibfnamefont {S.}~\bibnamefont {Leray}}, \ and\ \bibinfo {author}
  {\bibfnamefont {D.}~\bibnamefont {Mancusi}},\ }\href {\doibase
  10.1103/PhysRevC.87.014606} {\bibfield  {journal} {\bibinfo  {journal} {Phys.
  Rev. C}\ }\textbf {\bibinfo {volume} {87}},\ \bibinfo {pages} {014606}
  (\bibinfo {year} {2013})}\BibitemShut {NoStop}%
\bibitem [{\citenamefont {Ferrari}\ \emph
  {et~al.}(1996{\natexlab{a}})\citenamefont {Ferrari}, \citenamefont {Sala},
  \citenamefont {Ranft},\ and\ \citenamefont {Roesler}}]{Ferrari:1995cq}%
  \BibitemOpen
  \bibfield  {author} {\bibinfo {author} {\bibfnamefont {A.}~\bibnamefont
  {Ferrari}}, \bibinfo {author} {\bibfnamefont {P.~R.}\ \bibnamefont {Sala}},
  \bibinfo {author} {\bibfnamefont {J.}~\bibnamefont {Ranft}}, \ and\ \bibinfo
  {author} {\bibfnamefont {S.}~\bibnamefont {Roesler}},\ }\href {\doibase
  10.1007/s002880050119} {\bibfield  {journal} {\bibinfo  {journal} {Z. Phys.
  C}\ }\textbf {\bibinfo {volume} {70}},\ \bibinfo {pages} {413} (\bibinfo
  {year} {1996}{\natexlab{a}})},\ \Eprint
  {http://arxiv.org/abs/nucl-th/9509039} {arXiv:nucl-th/9509039} \BibitemShut
  {NoStop}%
\bibitem [{\citenamefont {Ferrari}\ \emph
  {et~al.}(1996{\natexlab{b}})\citenamefont {Ferrari}, \citenamefont {Sala},
  \citenamefont {Ranft},\ and\ \citenamefont {Roesler}}]{Ferrari:1996fv}%
  \BibitemOpen
  \bibfield  {author} {\bibinfo {author} {\bibfnamefont {A.}~\bibnamefont
  {Ferrari}}, \bibinfo {author} {\bibfnamefont {P.~R.}\ \bibnamefont {Sala}},
  \bibinfo {author} {\bibfnamefont {J.}~\bibnamefont {Ranft}}, \ and\ \bibinfo
  {author} {\bibfnamefont {S.}~\bibnamefont {Roesler}},\ }\href {\doibase
  10.1007/s002880050149} {\bibfield  {journal} {\bibinfo  {journal} {Z. Phys.
  C}\ }\textbf {\bibinfo {volume} {71}},\ \bibinfo {pages} {75} (\bibinfo
  {year} {1996}{\natexlab{b}})},\ \Eprint
  {http://arxiv.org/abs/nucl-th/9603010} {arXiv:nucl-th/9603010} \BibitemShut
  {NoStop}%
\bibitem [{\citenamefont {Ferrari}\ \emph {et~al.}(2005)\citenamefont
  {Ferrari}, \citenamefont {Sala}, \citenamefont {Fasso},\ and\ \citenamefont
  {Ranft}}]{Ferrari:2005zk}%
  \BibitemOpen
  \bibfield  {author} {\bibinfo {author} {\bibfnamefont {A.}~\bibnamefont
  {Ferrari}}, \bibinfo {author} {\bibfnamefont {P.~R.}\ \bibnamefont {Sala}},
  \bibinfo {author} {\bibfnamefont {A.}~\bibnamefont {Fasso}}, \ and\ \bibinfo
  {author} {\bibfnamefont {J.}~\bibnamefont {Ranft}},\ }\href {\doibase
  10.2172/877507} {\  (\bibinfo {year} {2005}),\ 10.2172/877507}\BibitemShut
  {NoStop}%
\bibitem [{\citenamefont {Ballarini}\ \emph {et~al.}(2024)\citenamefont
  {Ballarini} \emph {et~al.}}]{Ballarini:2024isa}%
  \BibitemOpen
  \bibfield  {author} {\bibinfo {author} {\bibfnamefont {F.}~\bibnamefont
  {Ballarini}} \emph {et~al.},\ }\href {\doibase 10.1051/epjn/2024015}
  {\bibfield  {journal} {\bibinfo  {journal} {EPJ Nuclear Sci. Technol.}\
  }\textbf {\bibinfo {volume} {10}},\ \bibinfo {pages} {16} (\bibinfo {year}
  {2024})}\BibitemShut {NoStop}%
\bibitem [{\citenamefont {Kelic}\ \emph {et~al.}(2009)\citenamefont {Kelic},
  \citenamefont {Ricciardi},\ and\ \citenamefont {Schmidt}}]{Kelic:2009yg}%
  \BibitemOpen
  \bibfield  {author} {\bibinfo {author} {\bibfnamefont {A.}~\bibnamefont
  {Kelic}}, \bibinfo {author} {\bibfnamefont {M.~V.}\ \bibnamefont
  {Ricciardi}}, \ and\ \bibinfo {author} {\bibfnamefont {K.-H.}\ \bibnamefont
  {Schmidt}}\ }(\bibinfo {year} {2009})\ \bibinfo {note} {"Joint ICTP-IAEA
  Advanced Workshop on Model Codes for Spallation Reactions"},\ \Eprint
  {http://arxiv.org/abs/0906.4193} {arXiv:0906.4193 [nucl-th]} \BibitemShut
  {NoStop}%
\bibitem [{\citenamefont {Haak}\ \emph {et~al.}(2023)\citenamefont {Haak} \emph
  {et~al.}}]{Haak:2023fbv}%
  \BibitemOpen
  \bibfield  {author} {\bibinfo {author} {\bibfnamefont {K.}~\bibnamefont
  {Haak}} \emph {et~al.},\ }\href {\doibase 10.1103/PhysRevC.108.034608}
  {\bibfield  {journal} {\bibinfo  {journal} {Phys. Rev. C}\ }\textbf {\bibinfo
  {volume} {108}},\ \bibinfo {pages} {034608} (\bibinfo {year}
  {2023})}\BibitemShut {NoStop}%
\bibitem [{\citenamefont {Ostroumov}\ \emph {et~al.}(2024)\citenamefont
  {Ostroumov} \emph {et~al.}}]{Ostroumov:2024kue}%
  \BibitemOpen
  \bibfield  {author} {\bibinfo {author} {\bibfnamefont {P.~N.}\ \bibnamefont
  {Ostroumov}} \emph {et~al.},\ }\href {\doibase
  10.1103/PhysRevAccelBeams.27.060101} {\bibfield  {journal} {\bibinfo
  {journal} {Phys. Rev. Accel. Beams}\ }\textbf {\bibinfo {volume} {27}},\
  \bibinfo {pages} {060101} (\bibinfo {year} {2024})},\ \bibinfo {note}
  {[Erratum: Phys.Rev.Accel.Beams 27, 089901 (2024)]}\BibitemShut {NoStop}%
\bibitem [{\citenamefont {Chang}\ \emph {et~al.}(2022)\citenamefont {Chang},
  \citenamefont {Aschenauer}, \citenamefont {Baker}, \citenamefont {Jentsch},
  \citenamefont {Lee}, \citenamefont {Tu}, \citenamefont {Yin},\ and\
  \citenamefont {Zheng}}]{Chang:2022hkt}%
  \BibitemOpen
  \bibfield  {author} {\bibinfo {author} {\bibfnamefont {W.}~\bibnamefont
  {Chang}}, \bibinfo {author} {\bibfnamefont {E.-C.}\ \bibnamefont
  {Aschenauer}}, \bibinfo {author} {\bibfnamefont {M.~D.}\ \bibnamefont
  {Baker}}, \bibinfo {author} {\bibfnamefont {A.}~\bibnamefont {Jentsch}},
  \bibinfo {author} {\bibfnamefont {J.-H.}\ \bibnamefont {Lee}}, \bibinfo
  {author} {\bibfnamefont {Z.}~\bibnamefont {Tu}}, \bibinfo {author}
  {\bibfnamefont {Z.}~\bibnamefont {Yin}}, \ and\ \bibinfo {author}
  {\bibfnamefont {L.}~\bibnamefont {Zheng}},\ }\href {\doibase
  10.1103/PhysRevD.106.012007} {\bibfield  {journal} {\bibinfo  {journal}
  {Phys. Rev. D}\ }\textbf {\bibinfo {volume} {106}},\ \bibinfo {pages}
  {012007} (\bibinfo {year} {2022})},\ \Eprint
  {http://arxiv.org/abs/2204.11998} {arXiv:2204.11998 [physics.comp-ph]}
  \BibitemShut {NoStop}%
\bibitem [{\citenamefont {Sjostrand}\ \emph {et~al.}(2006)\citenamefont
  {Sjostrand}, \citenamefont {Mrenna},\ and\ \citenamefont
  {Skands}}]{Sjostrand:2006za}%
  \BibitemOpen
  \bibfield  {author} {\bibinfo {author} {\bibfnamefont {T.}~\bibnamefont
  {Sjostrand}}, \bibinfo {author} {\bibfnamefont {S.}~\bibnamefont {Mrenna}}, \
  and\ \bibinfo {author} {\bibfnamefont {P.~Z.}\ \bibnamefont {Skands}},\
  }\href {\doibase 10.1088/1126-6708/2006/05/026} {\bibfield  {journal}
  {\bibinfo  {journal} {JHEP}\ }\textbf {\bibinfo {volume} {05}},\ \bibinfo
  {pages} {026} (\bibinfo {year} {2006})},\ \Eprint
  {http://arxiv.org/abs/hep-ph/0603175} {arXiv:hep-ph/0603175} \BibitemShut
  {NoStop}%
\bibitem [{\citenamefont {{Dupr\'e, Rapha\"el}}(2011)}]{Dupre:2011afa}%
  \BibitemOpen
  \bibfield  {author} {\bibinfo {author} {\bibnamefont {{Dupr\'e,
  Rapha\"el}}},\ }\href@noop {} {Ph.D. thesis},\ \bibinfo  {school} {Lyon, IPN}
  (\bibinfo {year} {2011})\BibitemShut {NoStop}%
\bibitem [{\citenamefont {Salgado}\ and\ \citenamefont
  {Wiedemann}(2003)}]{Salgado:2003gb}%
  \BibitemOpen
  \bibfield  {author} {\bibinfo {author} {\bibfnamefont {C.~A.}\ \bibnamefont
  {Salgado}}\ and\ \bibinfo {author} {\bibfnamefont {U.~A.}\ \bibnamefont
  {Wiedemann}},\ }\href {\doibase 10.1103/PhysRevD.68.014008} {\bibfield
  {journal} {\bibinfo  {journal} {Phys. Rev. D}\ }\textbf {\bibinfo {volume}
  {68}},\ \bibinfo {pages} {014008} (\bibinfo {year} {2003})},\ \Eprint
  {http://arxiv.org/abs/hep-ph/0302184} {arXiv:hep-ph/0302184} \BibitemShut
  {NoStop}%
\bibitem [{\citenamefont {Roesler}\ \emph {et~al.}(2000)\citenamefont
  {Roesler}, \citenamefont {Engel},\ and\ \citenamefont
  {Ranft}}]{Roesler:2000he}%
  \BibitemOpen
  \bibfield  {author} {\bibinfo {author} {\bibfnamefont {S.}~\bibnamefont
  {Roesler}}, \bibinfo {author} {\bibfnamefont {R.}~\bibnamefont {Engel}}, \
  and\ \bibinfo {author} {\bibfnamefont {J.}~\bibnamefont {Ranft}}\ }(\bibinfo
  {year} {2000})\ pp.\ \bibinfo {pages} {1033--1038},\ \Eprint
  {http://arxiv.org/abs/hep-ph/0012252} {arXiv:hep-ph/0012252} \BibitemShut
  {NoStop}%
\bibitem [{\citenamefont {Arleo}\ \emph {et~al.}(2019)\citenamefont {Arleo},
  \citenamefont {Na{\"\i}m},\ and\ \citenamefont {Platchkov}}]{Arleo:2018zjw}%
  \BibitemOpen
  \bibfield  {author} {\bibinfo {author} {\bibfnamefont {F.}~\bibnamefont
  {Arleo}}, \bibinfo {author} {\bibfnamefont {C.-J.}\ \bibnamefont
  {Na{\"\i}m}}, \ and\ \bibinfo {author} {\bibfnamefont {S.}~\bibnamefont
  {Platchkov}},\ }\href {\doibase 10.1007/JHEP01(2019)129} {\bibfield
  {journal} {\bibinfo  {journal} {JHEP}\ }\textbf {\bibinfo {volume} {01}},\
  \bibinfo {pages} {129} (\bibinfo {year} {2019})},\ \Eprint
  {http://arxiv.org/abs/1810.05120} {arXiv:1810.05120 [hep-ph]} \BibitemShut
  {NoStop}%
\bibitem [{\citenamefont {B\"ohlen}\ \emph {et~al.}(2014)\citenamefont
  {B\"ohlen}, \citenamefont {Cerutti}, \citenamefont {Chin}, \citenamefont
  {Fass\`o}, \citenamefont {Ferrari}, \citenamefont {Ortega}, \citenamefont
  {Mairani}, \citenamefont {Sala}, \citenamefont {Smirnov},\ and\ \citenamefont
  {Vlachoudis}}]{Bohlen:2014buj}%
  \BibitemOpen
  \bibfield  {author} {\bibinfo {author} {\bibfnamefont {T.~T.}\ \bibnamefont
  {B\"ohlen}}, \bibinfo {author} {\bibfnamefont {F.}~\bibnamefont {Cerutti}},
  \bibinfo {author} {\bibfnamefont {M.~P.~W.}\ \bibnamefont {Chin}}, \bibinfo
  {author} {\bibfnamefont {A.}~\bibnamefont {Fass\`o}}, \bibinfo {author}
  {\bibfnamefont {A.}~\bibnamefont {Ferrari}}, \bibinfo {author} {\bibfnamefont
  {P.~G.}\ \bibnamefont {Ortega}}, \bibinfo {author} {\bibfnamefont
  {A.}~\bibnamefont {Mairani}}, \bibinfo {author} {\bibfnamefont {P.~R.}\
  \bibnamefont {Sala}}, \bibinfo {author} {\bibfnamefont {G.}~\bibnamefont
  {Smirnov}}, \ and\ \bibinfo {author} {\bibfnamefont {V.}~\bibnamefont
  {Vlachoudis}},\ }\href {\doibase 10.1016/j.nds.2014.07.049} {\bibfield
  {journal} {\bibinfo  {journal} {Nucl. Data Sheets}\ }\textbf {\bibinfo
  {volume} {120}},\ \bibinfo {pages} {211} (\bibinfo {year}
  {2014})}\BibitemShut {NoStop}%
\bibitem [{\citenamefont {Battistoni}\ \emph {et~al.}(2015)\citenamefont
  {Battistoni} \emph {et~al.}}]{Battistoni:2015epi}%
  \BibitemOpen
  \bibfield  {author} {\bibinfo {author} {\bibfnamefont {G.}~\bibnamefont
  {Battistoni}} \emph {et~al.},\ }\href {\doibase
  10.1016/j.anucene.2014.11.007} {\bibfield  {journal} {\bibinfo  {journal}
  {Annals Nucl. Energy}\ }\textbf {\bibinfo {volume} {82}},\ \bibinfo {pages}
  {10} (\bibinfo {year} {2015})}\BibitemShut {NoStop}%
\bibitem [{\citenamefont {Whalley}\ \emph {et~al.}(2005)\citenamefont
  {Whalley}, \citenamefont {Bourilkov},\ and\ \citenamefont
  {Group}}]{Whalley:2005nh}%
  \BibitemOpen
  \bibfield  {author} {\bibinfo {author} {\bibfnamefont {M.~R.}\ \bibnamefont
  {Whalley}}, \bibinfo {author} {\bibfnamefont {D.}~\bibnamefont {Bourilkov}},
  \ and\ \bibinfo {author} {\bibfnamefont {R.~C.}\ \bibnamefont {Group}}\
  }(\bibinfo {year} {2005})\ pp.\ \bibinfo {pages} {575--581},\ \Eprint
  {http://arxiv.org/abs/hep-ph/0508110} {arXiv:hep-ph/0508110} \BibitemShut
  {NoStop}%
\bibitem [{\citenamefont {Miller}\ \emph {et~al.}(2007)\citenamefont {Miller},
  \citenamefont {Reygers}, \citenamefont {Sanders},\ and\ \citenamefont
  {Steinberg}}]{Miller:2007ri}%
  \BibitemOpen
  \bibfield  {author} {\bibinfo {author} {\bibfnamefont {M.~L.}\ \bibnamefont
  {Miller}}, \bibinfo {author} {\bibfnamefont {K.}~\bibnamefont {Reygers}},
  \bibinfo {author} {\bibfnamefont {S.~J.}\ \bibnamefont {Sanders}}, \ and\
  \bibinfo {author} {\bibfnamefont {P.}~\bibnamefont {Steinberg}},\ }\href
  {\doibase 10.1146/annurev.nucl.57.090506.123020} {\bibfield  {journal}
  {\bibinfo  {journal} {Ann. Rev. Nucl. Part. Sci.}\ }\textbf {\bibinfo
  {volume} {57}},\ \bibinfo {pages} {205} (\bibinfo {year} {2007})},\ \Eprint
  {http://arxiv.org/abs/nucl-ex/0701025} {arXiv:nucl-ex/0701025} \BibitemShut
  {NoStop}%
\bibitem [{ePI()}]{ePIC}%
  \BibitemOpen
  \href@noop {} {\enquote {\bibinfo {title} {The epic collaboration},}\
  }\bibinfo {howpublished} {\url{https://www.epic-eic.org/}}\BibitemShut
  {NoStop}%
\end{thebibliography}%
%------------------------------------------------------------------------

%------------------------------------------------------------------------
%------------------------------------------------------------------------
\end{document}